\documentclass[journal]{IEEEtran}
\usepackage{cite}
\ifCLASSINFOpdf
\usepackage[pdftex]{graphicx}
\else
\fi
\usepackage{amsmath}
\usepackage{algorithmic}
\usepackage{algorithm}
\usepackage{array}
\ifCLASSOPTIONcompsoc
\usepackage[caption=false,font=normalsize,labelfont=sf,textfont=sf]{subfig}
\else
\usepackage[caption=false,font=footnotesize]{subfig}
\fi
\newcommand\mypound{\protect\scalebox{1}{\protect\raisebox{0ex}{\#}}} % for a better \# symbol (Luan)

\usepackage[table]{xcolor} % Table color (Luan)
\usepackage{booktabs}

\usepackage{bm}
\usepackage{mathrsfs} %»¨Ìå hua ti zi mu     \mathscr{}
\usepackage{amsfonts,amssymb}  % ¿ÕÐÄ×Öĸ kong xin  \mathbb{}
\usepackage{amsopn}
\DeclareMathOperator*{\argmax}{argmax}                                          % argmax
\DeclareMathOperator*{\argmin}{argmin} 

\usepackage{lipsum} % random words

\begin{document}
\title{AoA-aware Probabilistic Indoor Location Fingerprinting using Channel State Information}

\author{Luan~Chen,
        Iness~Ahriz,
        and~Didier~Le~Ruyet,~\IEEEmembership{Senior Member,~IEEE}% <-this % stops a space
%\thanks{Manuscript received December, 2019; revised 2020. This work was supported in part by China Scholarship Council (Beijing, China) under Grant CSC201606270203, and the National Natural Science Foundation of China under Grant No. 61771014. (Corresponding author: Luan Chen)}
\thanks{The authors are with CEDRIC/LAETITIA Laboratory, Conservatoire National des Arts et M\'{e}tiers (CNAM), 75003 Paris, France.}% <-this % stops a space
\thanks{Luan Chen is also with School of Electronic Information, Wuhan University, 430072 Wuhan, China. (e-mail: luan.chen@whu.edu.cn).}}% <-this % stops a space


\markboth{Journal of \LaTeX\ Class Files,~Vol.~14, No.~8, August~2015}{Chen \MakeLowercase{\textit{et al.}}: AoA-aware Probabilistic Indoor Location Fingerprinting using Channel State Information}



\maketitle

\begin{abstract}
With expeditious development of wireless communications, location fingerprinting (LF) has nurtured considerable indoor location based services (ILBSs) in the field of Internet of Things (IoT). For most pattern-matching based LF solutions, previous works either appeal to the simple received signal strength (RSS), which suffers from dramatic performance degradation due to sophisticated environmental dynamics, or rely on the fine-grained physical layer channel state information (CSI), whose intricate structure leads to an increased computational complexity. Meanwhile, the harsh indoor environment can also breed similar radio signatures among certain predefined reference points (RPs), which may be randomly distributed in the area of interest, thus mightily tampering the location mapping accuracy. To work out these dilemmas, during the offline site survey, we first adopt autoregressive (AR) modeling entropy of CSI amplitude as location fingerprint, which shares the structural simplicity of RSS while reserving the most location-specific statistical channel information. Moreover, an additional angle of arrival (AoA) fingerprint can be accurately retrieved from CSI phase through an enhanced subspace based algorithm, which serves to further eliminate the error-prone RP candidates. In the online phase, by exploiting both CSI amplitude and phase information, a novel bivariate kernel regression scheme is proposed to precisely infer the target's location. Results from extensive indoor experiments validate the superior localization performance of our proposed system over previous approaches.

%intricate  harsh  dynamics thriving  yield   actualize     empower   tamper  cumbersome  mightily   nurture  cutting-edge  thriving cultivate  fertilize individually  In the sequel, shed light on  dedicated  underperforming  amenable  encapsulate  ramp up  cope with  workaround 
%\lipsum[4-5]
\end{abstract}

\begin{IEEEkeywords}
Internet of things, location fingerprinting, channel state information, entropy, angle of arrival.
\end{IEEEkeywords}

% \fi
%\IEEEpeerreviewmaketitle

\section{Introduction}\label{sec:introduction}
\IEEEPARstart{W}{ith} the wide-scale proliferation of wireless communication and ubiquitous computing, indoor location based service (ILBS) has emerged as a key enabler for myriad cutting-edge applications in the domain of Internet of Things (IoT) \cite{kuutti2018survey,zafari2019survey,he2015wi,dardari2015indoor}. Such examples include autonomous navigation for visually-impaired individuals, logistic monitoring in smart warehouse, proximity marketing at commercial hub, intruder tracking in sensitive facilities, etc.. Due to the pervasive availability and the low-cost deployment, Wi-Fi based location awareness stands out as one of the most appealing solutions in lieu of many other wireless techniques, e.g., Bluetooth \cite{liu2013face}, radio frequency identification (RFID) \cite{ni2003landmarc}, ultrasound \cite{ward1997new}, infrared \cite{want1992active}, visible light \cite{pathak2015visible} and so forth. In general, conventional Wi-Fi based indoor positioning systems (IPS) either adopt geometric mapping approach or resort to location fingerprinting (LF) \cite{yang2013rssi}. For geometric mapping, spatial properties like time of flight (ToF) \cite{tadayon2019decimeter} or angle of arrival (AoA) \cite{kotaru2015spotfi} are first derived from physical communication signals. Target's position determination is then conducted through geometric algorithms (e.g., trilateration or triangulation), which however heavily rely on the line-of-sight (LoS) condition. This makes geometric mapping less eligible for the sophisticated indoor environment with rich hindrances and room partitions. As an emerging alternative for indoor positioning, location fingerprinting benefits from a pattern-matching mechanism, which comprises offline training phase and online location estimation phase. Specifically, in the offline phase, wireless signatures are collected at a set of geo-tagged reference points (RPs) in the area of interest to construct the fingerprint database (a.k.a. radio map). During the online phase, the measured signature at an unknown position is matched with the offline radio map to return the best-fitted location estimation. 

Although the mainstream Wi-Fi fingerprinting systems take the simple received signal strength (RSS) as the indicator of medium access control (MAC) layer's link quality, it suffers dramatic performance degradation due to small-scale multipath fading and temporal dynamics indoors. In virtue of the break-through technology of multiple-input multiple-output orthogonal frequency division multiplexing (MIMO-OFDM) in IEEE 802.11 n/ac standard, the fine-grained physical (PHY) layer channel state information (CSI) is capable of depicting the channel characterization for each transmit-receive (TX-RX) antenna pair on the level of multiple orthogonal subcarriers \cite{yang2013rssi,wu2012csi}. Different from coarse-grained RSS, CSI captures the amplitude attenuation and phase shift of every subcarrier and thus can serve as a preferable geo-signature to bring richer location-specific information for numerous Wi-Fi fingerprinting systems. 
%In addition, the current accessibility of CSI, which can be partially extracted from many commercial off-the-shelf network interface cards (NICs), advances its great popularity among numerous Wi-Fi fingerprinting systems.

In principle, Wi-Fi fingerprinting algorithms can be categorized into deterministic and probabilistic ones \cite{he2015wi}. Deterministic approaches enjoy the easy implementation but fail to fully exploit environmental fluctuations, which consequently renders the location estimation error-prone. In contrast, probabilistic methods embrace the channel variation by inferring a signal distribution based statistical model, thus obtaining more robust and accurate positioning performance than its deterministic adversary. Nevertheless, there still exists three underlying challenges for probabilistic Wi-Fi fingerprinting systems: (i) The accurate approximation of probability distribution function (PDF) is largely driven by massive storage of signal measurements \cite{alsindi2014empirical}, which in turn brings huge system burden and computational requirement. (ii) Most probabilistic location-aware solutions are well established on the assumption of Gaussian-distributed measurements \cite{youssef2005horus,xiao2012fifs}. However, due to the complex nature of indoor environment and the imperfection of wireless devices, some practical measurements appear to be non-Gaussian distributed or even do not fit any known distribution \cite{sen2012you,zhou2013omnidirectional,mirowski2011kl}. This then complicates the fingerprinting process and incurs severe ambiguity for location estimation. (iii) When it comes to multivariate fingerprint structure (e.g. multi-subcarrier CSI), traditional probabilistic methods turn powerless since existing statistical tools only work for measurements with identifiable distributions \cite{chen2018probabilistic}. Accordingly, it would be highly desirable for a fingerprint which shares the simplicity of RSS (scalar) and meanwhile conserves rich statistical location-specific information. 

To address the aforementioned substantial challenges, in this paper, we resort to autoregressive (AR) modeling based Shannon entropy metric \cite{bercher2000estimating}, which equals a direct transformation from the original PDF of CSI amplitudes. Unlike traditional data-adaptive histogram estimator which entails a slow convergence rate, AR modeling approach provides a feasible parametric workaround to accurately infer the PDF in the form of power spectral density (PSD) \cite{kay1998model}. Despite its structural simplicity, this novel entropy fingerprint embodies the whole statistical information of CSI amplitudes. Through extensive experiments conducted in realistic testbeds, we demonstrate that our proposed AR entropy metric outperforms its original CSI or RSS fingerprint \cite{luan2019entropy}. However, since CSI phases of one subcarrier are generally uniformly distributed \cite{molisch2012wireless}, this quantifies each RP location with an equally maximized entropy value (a.k.a. Gibbs' inequality), thereby hampering the location distinction to a great extend. How to properly exploit CSI phase information in our entropy-based location fingerprinting system still remains open. 

\begin{figure}[!t]
	\centering
	\includegraphics[width=1\linewidth]{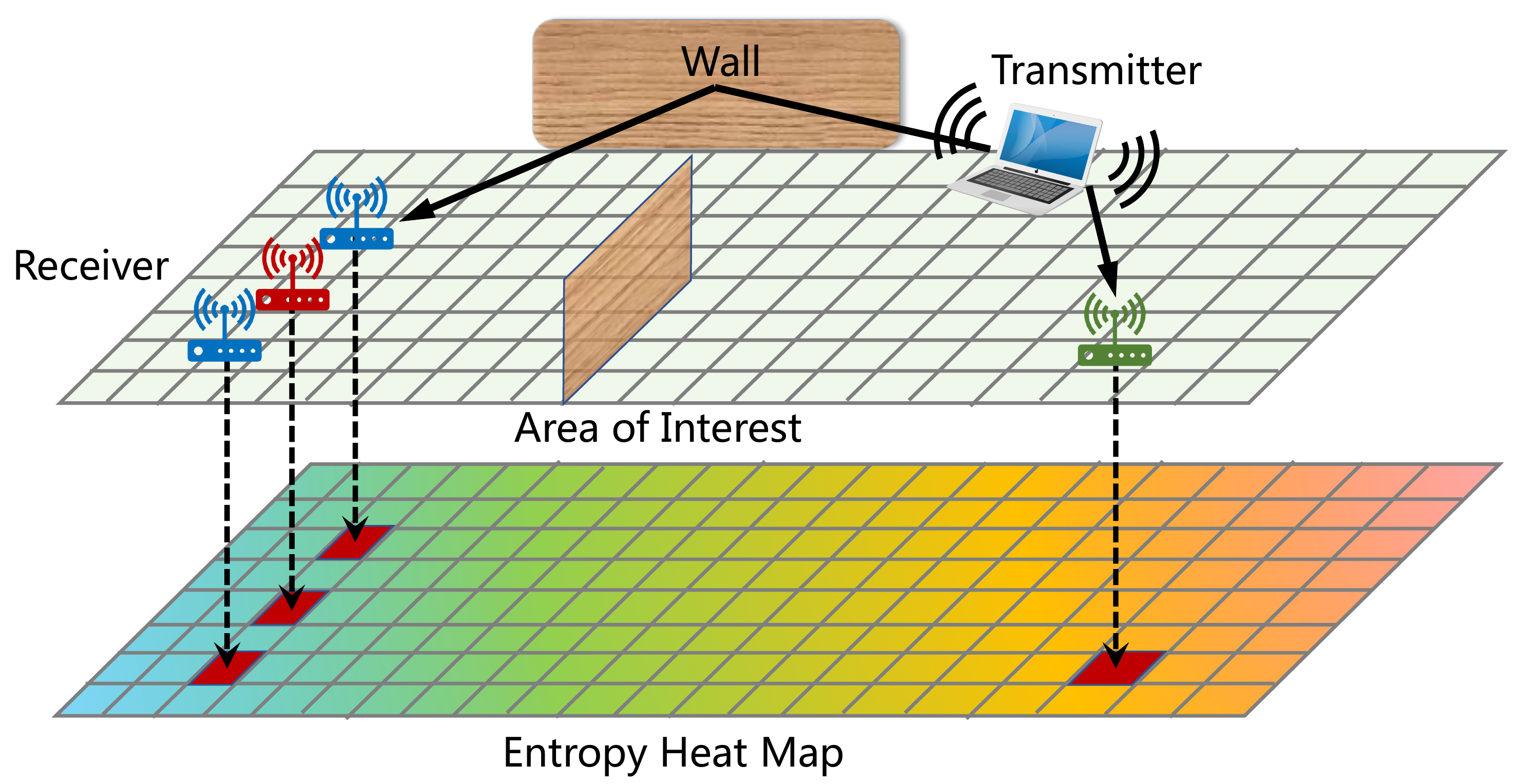}
	\caption{An illustrative example of the system mechanism}
	\label{ruleout}
\end{figure}

Inspired by the recent advancement of phased array signal processing \cite{kotaru2015spotfi}, leveraging AoA as supplementary fingerprint enables us to revisit CSI phase exploitation with a fresh horizon. As illustrated in Fig. \ref{ruleout}, our AoA embedded solution adopts the methodological concept of the well-known k-nearest neighbors (kNN) and unveils two heuristic insights: (i) For some offline surveying receivers at the corresponding RP positions, whether they are in the vicinity (blue ones) of the online receiver (red one) or in the distance (green one), their CSI measurements may share the similar entropy values. (ii) These neighboring receivers also record the similar AoAs from parallel incident paths with this online receiver, whether it is for direct paths in LoS scenario or reflected paths in NLoS condition. Hence, the remote receiver can be selectively ruled out in accordance with the distinct AoA difference, which further improves the location estimation accuracy. %With the virtual antenna extension and careful phase calibration, the enhanced multiple signal classification (MUSIC) algorithm is able to achieve high resolution of AoA estimation, which further contributes to the boost of the fingerprinting accuracy. 

On this basis, we design AngLoc, an AoA-aware probabilistic indoor localization system using commercial off-the-shelf Wi-Fi device. To remove the noisy component from the raw CSI measurements, we first introduce a power-based tap filtering scheme to preserve the most informative CSI signatures. For the purpose of precise AoA estimation, a set of phase calibration techniques are then employed to mitigate dramatic phase drifts. Subsequently, for the offline radio map construction, the pre-processed CSIs are simultaneously fed to two independent fingerprint generators, namely AR modeling based entropy estimator for CSI amplitude and the enhanced AoA-ToF estimator driven by joint angle and delay estimation multiple signal classification (JADE-MUSIC) algorithm \cite{vanderveen1997joint}. It is worth noting that ToF is utilized here to create measurable phase shift across subcarriers, by which realizes virtual antenna extension to overcome the antenna number restriction for classical MUSIC algorithm \cite{kotaru2015spotfi}. The other trick of ToF here is to identify the first incoming path (not necessarily the direct path) as the angular fingerprint benchmark, which serves to guarantee similar AoA recordings around closely-spaced RPs. Moreover, in the online phase, due to the simple structure of the radio map, the succinct Manhattan distance and Euclidean distance can be fully competent as the similarity metrics for AR entropy and AoA fingerprints, respectively. Afterwards, we propose an optimal bivariate kernel regression scheme to accurately infer the target's physical location. The entire experiments are conducted on the lightweight HummingBoard platform, which tremendously facilitates the time-consuming and labor-intensive fingerprinting implementation. Experimental results validate the superior performance of our proposed system over previous location fingerprinting approaches. 
%The estimated ToF values are not the absolute ToFs. Our main purpose is to achieve AoA diversity. Accurate ToF estimation is beyond the scope of this article.

In a nutshell, the main contributions of this paper can be laid out below:
\begin{enumerate}
\item[$\bullet$] As far as we are aware of, this is the first work to constructively incorporate angular signature in CSI entropy-based indoor location fingerprinting system, fertilizing the opportunity to achieve a decimeter-level accuracy. %fostering
%We propose a novel AoA-aware fingerprinting mechanism which exploits statistical AR entropy using CSI amplitude, then further refine the estimation through AoA based RP sifting. The incorporation of AR entropy and AoA sufficiently CSI amplitude and phase information, boosting the performance 
%As far as we are aware of, this is the first work to incorporate CSI amplitude based AR entropy and CSI phase based AoA-ToF as location fingerprints to achieve decimeter-level estimation accuracy. 

\item[$\bullet$] We propose a power-based tap-filtering program alongwith several phase calibration pre-processing techniques to effectively mitigate CSI noisy component and sanitize CSI phase errors, respectively.

\item[$\bullet$] We design a feasible bivariate kernel regression scheme for the online location estimation stage, which organically combines the weighting factors for both amplitude-based entropy and phase-based AoA fingerprints.

\item[$\bullet$] We build and implement extensive experiments on the lightweight Hummingboard device for different testbeds. In addition to the superior performance, our mobile prototype remarkably enhances the fingerprinting efficiency.

\end{enumerate}

The remainder of this paper is organized as follows. In Section \ref{sec:relatedwork}, we review the state-of-the-art related works. The relevant preliminaries are introduced in Section \ref{sec:preliminaries}. The overall architecture design of our proposed system is elaborated in Section \ref{sec:system}. We provide experimental results and analyses in Section \ref{sec:experiment} and dig some insightful perspectives in Section \ref{sec:discussion}. Conclusions are drawn in Section \ref{sec:conclusion}.

%%%%%%%%%%%%%%%%%%%%%%%%%%%%%%%%%%%%%%%%%%%%%%%%%%%%%%%%%%%%%%%%%%%%%%%%%%%%%%%%%%%%%%%%%%%%%%%%%%%%%%%%%%%
\section{Related Work}\label{sec:relatedwork}
The popularization of mobile computing triggers a thriving trend in the domain of wireless indoor location determination. The general localization approaches fall into two categories: geometrical mapping and fingerprinting.

\textbf{\textit{Geometric mapping based techniques}}:
The geometric modeling of the RF propagation is fundamental to the ranging or direction based positioning systems. Wu et al. explored the frequency diversity of PHY layer CSI information to refine distance estimation and pinpoint the target's location through trilateration in FILA system \cite{wu2012fila}, which achieved median accuracy of 1.2 m in the multi-room environment. Alternatively, ArrayTrack \cite{xiong2013arraytrack} embraced the trend of MIMO technology and exploited increased number of antennas at commodity access points (APs) to obtain high-resolution AoAs, which were further aggregated to infer the client location within 23 centimeters median accuracy. Unlike ArrayTrack which requires dedicated hardware modifications, Kotaru et al. designed SpotFi \cite{kotaru2015spotfi}, an accurate indoor localization system capable of identifying direct path AoAs with only three physical RX antennas. Moreover, after incorporating the observed RSS information for an optimization processing, SpotFi was able to achieve the median accuracy of 40 cm. More recently, the researchers of Chronos \cite{vasisht2016decimeter} leveraged a novel Chinese remainder theorem based algorithm to compute sub-nanosecond ToF with a single Wi-Fi access point. This distance-related metric was then formulated into a quadratic optimization problem for accurately locating clients within tens of centimeters.   

\textbf{\textit{Fingerprinting based techniques}}:
Regardless of measurements' geometric relation, the pattern-matching based fingerprinting techniques have attracted a large body of research interests for the last decades. Pioneering works such as RADAR \cite{bahl2000radar} carried out comprehensive site surveys for the first time and generated the RSS based fingerprint radio map. Subsequently, the deterministic kNN algorithm was utilized to determine the target's location with an average precision of 3 meters. Contrastively, in Horus system \cite{youssef2005horus}, Youssef et al. resorted to the Bayes based probabilistic method and a joint clustering algorithm to achieve an accuracy improvement of 2.1 m, which outperformed RADAR even with less computational complexity. Aside from FILA, the authors of FIFS \cite{xiao2012fifs} also explored the spatial and frequency diversity of CSI for Wi-Fi fingerprinting localization. Additionally, FIFS took the power summation for all independent subcarriers as location fingerprint and adopted maximum a posteriori (MAP) approach to yield an improved performance compared with RSS based Horus system. Meanwhile, for PinLoc \cite{sen2012you}, the whole location-aware platform was established on a set of $1 m \times 1 m$ spots. The underlying observation of PinLoc was that the CSIs on a single subcarrier were illustrated to be clustered distributed on the complex plane. The Gaussian mixture distribution was then introduced to properly model the channel measurements for the purpose of accurate localization. Experimental result validated PinLoc's impressive performance with an $89 \%$ mean accuracy for 100 spots. Wang et al. designed DeepFi \cite{wang2016csi}, a deep learning based indoor location fingerprinting system using CSI amplitude information. In the offline phase, DeepFi enabled a deep network to train all the weights as location fingerprints, and harnessed the radial basis function (RBF) based probabilistic scheme to accomplish the position estimation in the online phase. It outperformed FIFS system with $20 \%$ accuracy improvement. Recently, authors in \cite{luan2019entropy} proposed EntLoc, an AR entropy based indoor location fingerprinting system using CSI amplitude information. Compared with PinLoc, the EntLoc's structural simplicity and the strong location-dependency of AR entropy based CSI fingerprint remarkably enhanced the location fingerprinting performance with an average precision improvement of $27.3 \%$.

%%%%%%%%%%%%%%%%%%%%%%%%%%%%%%%%%%%%%%%%%%%%%%%%%%%%%%%%%%%%%%%%%%%%%%%%%%%%%%%%%%%%%%%%%%%%%%%%%%%%%%%%%%%
\section{Preliminaries}\label{sec:preliminaries}
In this section, we present some relevant technical backgrounds on the MIMO-OFDM mechanism and the channel state information. 
\begin{figure}[!t]
	\centering
	\includegraphics[width=1.05\linewidth]{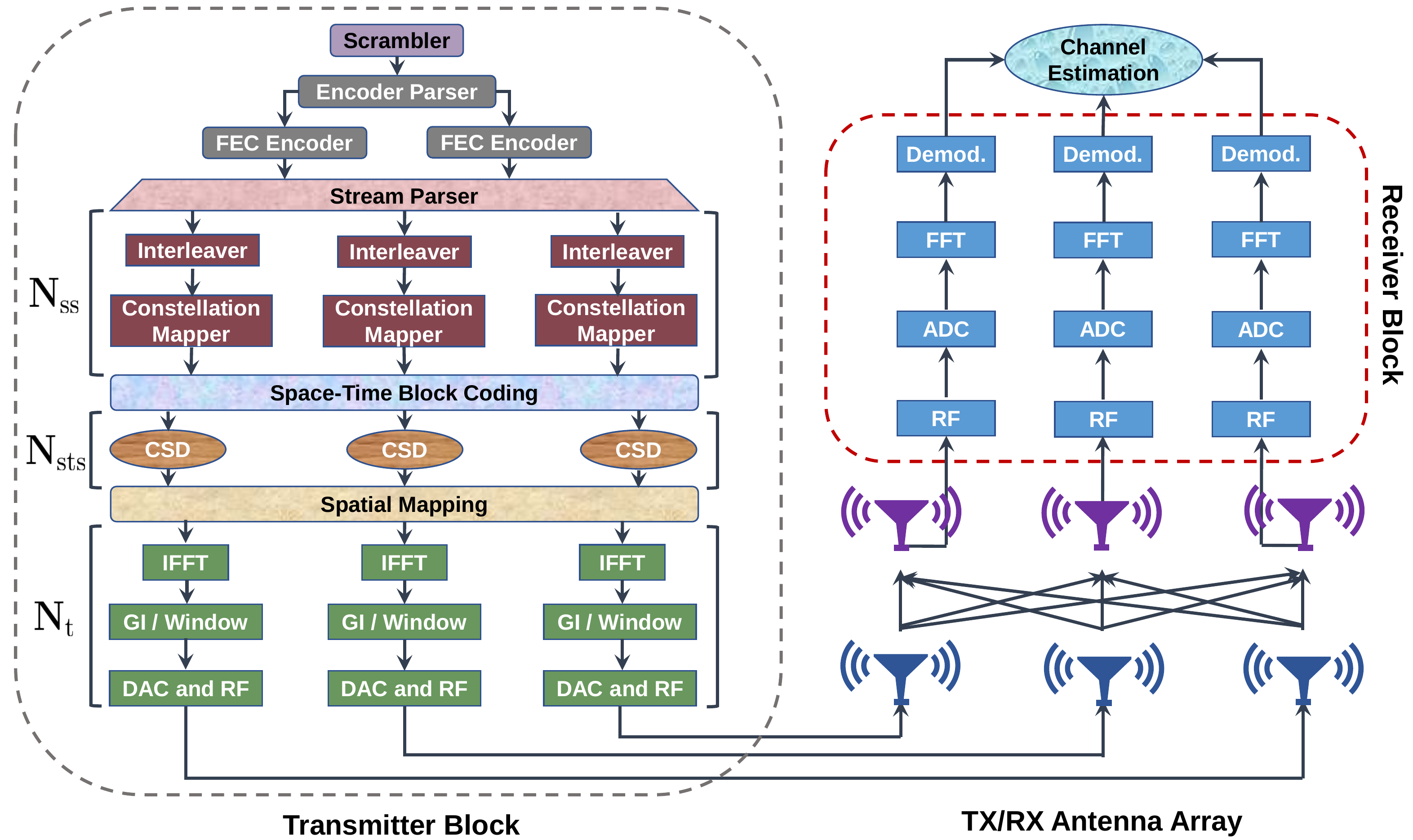}
	\caption{MIMO-OFDM transceiver architecture}
	\label{mimoofdm}
\end{figure}

\subsection{MIMO-OFDM Mechanism} \label{subsec:mimo}
Fig. \ref{mimoofdm} depicts the holistic structure of the end-to-end MIMO-OFDM wireless transceiver for IEEE 802.11 n/ac. It consists of two major functionality blocks: signal transmitter block and RF receiver block. In the transmitter block, the scrambler and forward error correction (FEC) encoder first convert the input data into high-rate bit stream(s). The stream parser is then applied on bit stream(s) to generate $N_{ss}$ spatial streams, whose number is determined by the parameter of modulation and coding scheme (MCS). After interleaving and constellation mapping (e.g. QAM), $N_{ss}$ spatial streams are modulated as stream of symbols, which may be spread into $N_{sts}$ space-time streams in the sequel when space-time block coding (STBC) is used. Next, a mechanism named cyclic shift diversity (CSD) is applied to insert cyclic shifts into space-time streams, thus creating extra frequency diversity to avoid unintentional beamforming. Spatial mapping then proceeds to map fewer number of $N_{sts}$ space-time streams into larger number of $N_t$ transmit chains through spatial mapping matrix (SMM). Afterwards, the frequency domain samples are converted into time domain ones by the inverse fast Fourier transform (IFFT). The RF signals are then simultaneously sent from all transmit antennas after the insertion of guard interval (GI), windowing operation and digital-to-analog converting (DAC).   % binary convolutional codes (BCC) There may be 1 or 2 FEC encoders when BCC encoding is used

In reverse, upon receiving the signals, the receiver block first samples them and digitizes them through analog-to-digital converters (ADCs). Subsequently, a forward FFT procedure is conducted to convert the data samples back to the frequency domain. The desired channel estimation then becomes achievable after the signal demodulation process.

\subsection{Channel State Information}
In wireless communication systems, the signal receiver operates channel estimation by virtue of channel sounding mechanism. Specifically, for the packet-based IEEE 802.11n system, the transmitter sends training sequences, including high throughput long training fields (HT-LTF) in the preamble. Once receiver detects the starting position of the first HT-LTF, it commences to derive channel state information immediately. As aforementioned in Section \ref{sec:introduction}, CSI portrays the PHY layer channel properties in the frequency domain and reveals the combined effects of signal multipath propagation including amplitude attenuation and phase shift. The channel frequency response (CFR) is represented by each CSI entry. It can be expressed by
\begin{equation}
H(f)=| H(f) |  e^{j \angle H(f)}
\end{equation}
where $H(f)$ is the complex value of CFR at the subcarrier with central frequency of $f$. $| H(f) |$ and $\angle H(f)$ denote its amplitude and phase, respectively.

In order to fully characterize the indoor multipaths, the time domain counterpart of CFR, also termed as channel impulse response (CIR), is able to model the wireless propagation channel as a temporal linear filter. Mathematically, it can be denoted as 
\begin{equation}
h(\tau)=\sum_{i=1}^{L} \alpha _i e^{-j \varphi _i} \delta(\tau-\tau _i )
\end{equation}
where $\alpha _i$, $\varphi _i$ and $\tau _i$ are the amplitude, phase and time delay spread of the $i^{th}$ path, respectively. $L$ is the total number of multipaths and $\delta(\cdot)$ is the Dirac delta function.

In practice, it is worth mentioning that, all of our experiments are implemented on the basis of Linux CSI tool \cite{Halperin_csitool}, whose off-the-shelf Intel 5300 network interface card (NIC) reports 30 out of 56 OFDM subcarriers for 20 MHz bandwidth CFR. After applying IFFT on the recorded CFR, we can acquire the time domain CIR with an equivalent number of 30 channel filter taps.

%%%%%%%%%%%%%%%%%%%%%%%%%%%%%%%%%%%%%%%%%%%%%%%%%%%%%%%%%%%%%%%%%%%%%%%%%%%%%%%%%%%%%%%%%%%%%%%%%%%%%%%%%%%
\section{System Design}\label{sec:system}
In this section, we lay out the detailed design of our proposed fingerprint localization system. 

\subsection{Overview}
\begin{figure}[!t]
	\centering
	\includegraphics[width=1\linewidth,height=1.8in]{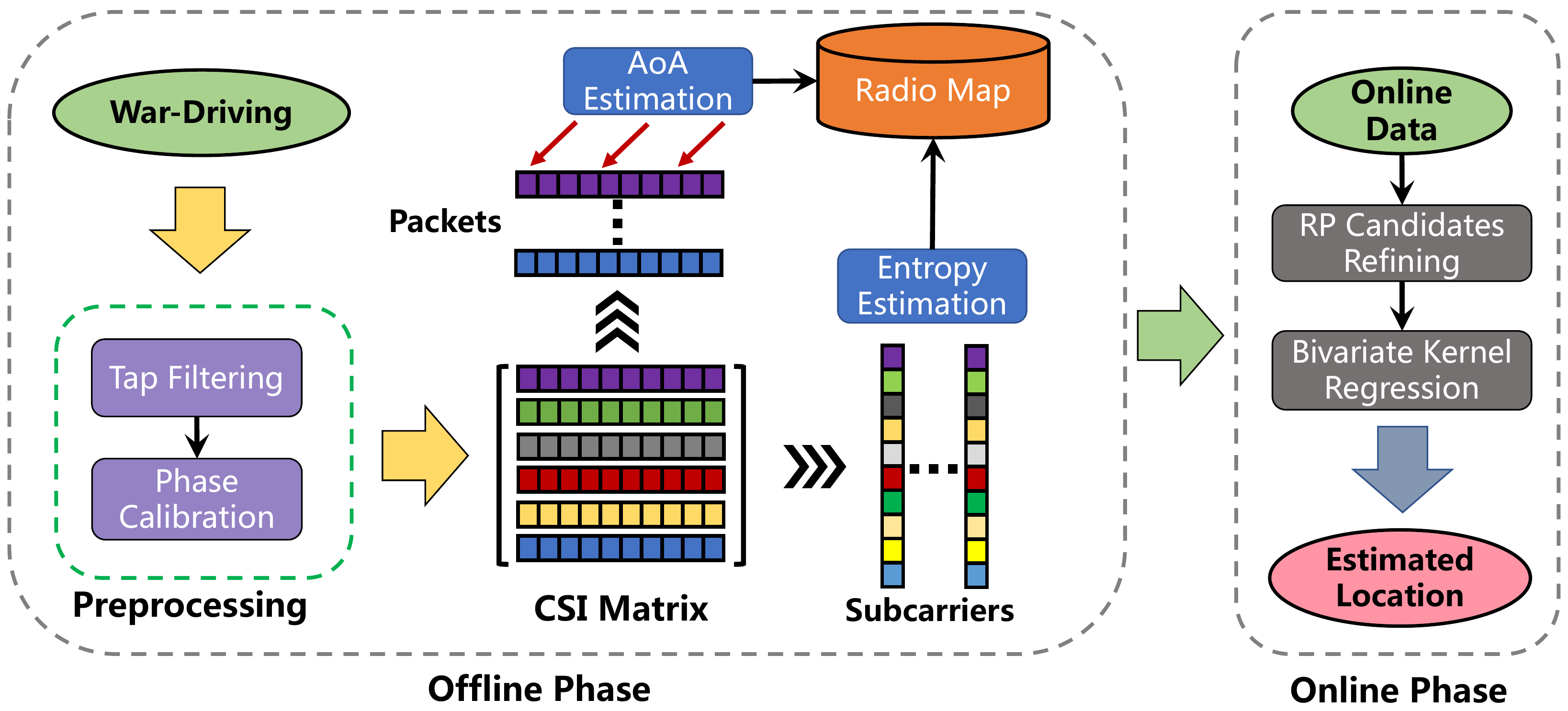}
	\caption{The overall AngLoc system architecture}
	\label{flowchart}
\end{figure}

%\IEEEpubidadjcol
As illustrated in Fig. \ref{flowchart}, the overall architecture of our proposed AngLoc system has a block-wise design. To be specific, in the offline radio map construction block, once recording the raw CSI measurements through war-driving, we first introduce a tap filtering scheme to extract the most informative location-specific component from noisy CSIs. For the purpose of accurate AoA estimation, several phase calibration techniques are then leveraged to compensate the corresponding phase offsets, which exist in prevalent commodity WiFi devices. Subsequently, for CSI amplitudes, we statistically model them as the simply structural AR entropy metric. The JADE-MUSIC algorithm is then adopted for CSI phases to infer the angular estimates. Hence, the entire offline database can be fully embodied by the integration of entropy and AoA fingerprints, making full use of both CSI amplitude and phase information. For the online location estimation block, when a mobile target enters the area of interest, it executes the same pre-processing protocols to obtain the entropy and AoA estimates. The following location estimation task then consists of two major steps. (i) The online entropy vector is first matched with offline database to find the most likely candidates from nearest RP positions. (ii) Among these candidate locations, a novel bivariate kernel regression scheme is proposed to further narrow down the number of error-prone RPs, thus tackling the target's location determination with an improved accuracy.

In the sequel, we will take an in-depth structural dissection for each block of our proposed AngLoc system.

\subsection{Problem Formulation}
First of all, we start to elaborate our location fingerprinting methodology with the presentation of the problem formulation. In the offline phase, $M$ reference points are predefined and properly marked in the area of interest. The coordinate of each RP location can be denoted as $\bm{\ell}_{m}=(x_{m}, y_{m})$, where $x_m$ and $y_m$ are the respective X- and Y-coordinate of the $m^{th}$ RP, $m \in [1,M]$. Considering that we have $S$ access points as signal transmitters, each of which has $N_t$ TX antennas. One mobile user equipped with $N_r$ RX antennas is regarded as the RF receiver. Thus each TX-RX antenna pair is capable of generating up to $N_t \cdot N_r$ radio links. As for the channel state information, each CSI packet shares the same number of $K$ OFDM subcarriers. So the dimensionality of one CSI packet measured at one RP location from a single AP can be expressed by $R=N_t \cdot N_r \cdot K$. Moreover, at each RP position, we propose to concatenate CSI packets from $S$ available APs to form the raw CSI signature, whose total dimensionality then extends to be $S \cdot R$. Mathematically, this offline radio signature measured at the $m^{th}$ RP location from all $S$ APs is given by the set $\bm{\mathcal{H}}_m=\{ {\mathbf{H}}^1_m, \dots, {\mathbf{H}}^s_m, \dots, {\mathbf{H}}^S_m \}, s \in [1,S]$. Specifically, ${\mathbf{H}}^s_m \in \mathbb{C}^{N \times R}$ contains $N$ consecutive $1 \times R$ dimensional CSI samples which are adequately acquired at the RP location $\bm{\ell}_{m}$ from the $s^{th}$ AP. This CSI matrix can be presented by the following equation.
\begin{equation} \label{equ:cfrmatrix}
\begin{small}
\mathbf{H}_m^s=
\begin{bmatrix} 
H_m^s (1,1) &  \cdots & H_m^s (1,r) & \cdots & H_m^s (1,R) \\
\vdots      &  \ddots & \vdots      & \ddots & \vdots      \\
H_m^s (n,1) &  \cdots & H_m^s (n,r) & \cdots & H_m^s (n,R) \\
\vdots      &  \ddots & \vdots      & \ddots & \vdots      \\
H_m^s (N,1) &  \cdots & H_m^s (N,r) & \cdots & H_m^s (N,R) 
\end{bmatrix}
\end{small}
\end{equation}
where $n \in[1, N]$ and $r \in [1, R]$.

During the online stage, the mobile user at an unknown position $\bm{\ell}_{o}=(x_{o}, y_{o})$ records the same structural CSI matrix from the $s^{th}$ AP. We denote this matrix as ${\mathbf{G}}_{o}^s$, which also shares the same dimension with ${\mathbf{H}}_m^s$. Likewise, the online measured CSI signature at the location $\bm{\ell}_{o}$ can be expressed by the set $\bm{\mathcal{G}}_{o}=\{ {\mathbf{G}}^1_{o}, \dots, {\mathbf{G}}^s_{o}, \dots, {\mathbf{G}}^S_{o} \}$. Accordingly, the mobile user's location can be then estimated as $\widehat{\bm{\ell}}_o=(\widehat{x}_o, \widehat{y}_o)$ by exploiting these online CSIs and the stored offline database.

%%%%%%%%%%%%%%%%%%%%%%%%%%%%%%%%%%%%%%%%%%%%%%%%%%%%%%%%%%%%%%%%%%%%%%%%%%%%%%%%%%%%%%%%%%%%%%%%%%%%%%%%%%%%
\subsection{CSI Pre-Processing}
In this part, we focus on some technical details of CSI pre-processing techniques which serve as the precondition to attain superior localization performance.

\subsubsection{CSI Noise Removal}
Due to the bandwidth limitation for the existing Wi-Fi networks, CSI fingerprint based IPS cannot resolve enough multipath components in the environment, which may incur severe ambiguity for location fingerprinting \cite{chen2016achieving}. For instance, from a commodity Wi-Fi with 20 MHz bandwidth, the time resolution of CIR can only reach 1/20MHz = 50 ns. Since typical indoor maximum excess delay is smaller than 500 ns, given a time resolution of 50 ns, just the first 10 out of the 30 accessible CIR time taps are relevant to multipath propagation \cite{zhou2015wifi}. In other words, the remaining 20 taps are less related for localization purpose. In addition, when exposed in a low signal-to-noise ratio (SNR) scenario, the receiver's additive white Gaussian noise at these time taps will even make the accuracy worse.

Therefore, a reasonable number of CIR taps should be carefully selected for the sake of precise positioning. On this basis, we design a power-based tap filtering scheme to preserve the most informative location dependency. Specifically, for the simplicity of expression, we define an individual raw CFR signature as $\mathbf{H} \in \mathbb{C}^{1 \times K}$. Through IFFT, we first convert $\mathbf{H}$ into the same dimensional CIR vector $\mathbf{h}$. Next, we calculate the average channel power of each tap, denoted by $\textbf{U}=(u_1, \dots, u_k, \dots, u_K), k \in [1,K]$, where $u_k=| \textrm{h}_k|^2$ and $\textrm{h}_k$ indicates the $k^{th}$ complex tap value of the CIR vector. Then, a cumulative contribution rate of the first $k$ taps is defined as
\begin{equation}
C_k=\sum_{i=1}^{k} u_i \bigg/ \sum_{i=1}^{K} u_i. 
\end{equation} 
If the cumulative contribution rate of the first $T$ taps, i.e., $C_T$, is greater than a predefined threshold $C$, we then apply a simple rectangular window with length $T$ to truncate the other $(K-T)$ taps. At last, FFT is further employed on the filtered CIR to yield a smoothed version of CFR \cite{luan2019entropy}.

\subsubsection{CSI Phase Calibration}
Due to the inherent OFDM baseband operations and the hardware's imperfect signal processing, the CSI obtained from the commodity Wi-Fi devices is distorted with various errors \cite{tadayon2019decimeter,ma2019wifi,zhuo2017perceiving}, rendering the accurate AoA and ToF estimation much more challenging. For a transmission chain, the phase measurement $\angle \widehat{H}_{f_k}$ for subcarrier $k$ with carrier frequency $f_k$ can be presented as
\begin{equation}
%\begin{split}
\angle \widehat{H}_{f_k}=
\angle H_{f_k} + 2 \pi f_\delta k  (\zeta_\mathbf{csd} + \xi_\mathbf{sfo} )  + \varphi_\mathbf{sto}
 + \varphi_\mathbf{cfo} + \varphi_\mathbf{cpo} +Z
%\end{split}
\end{equation}
where $\angle H_{f_k}$ denotes the true phase from wireless propagation. $f_\delta$ is the OFDM subcarrier spacing. $\zeta_\mathbf{csd}$, $\xi_\mathbf{sfo}$, $\varphi_\mathbf{sto}$, $\varphi_\mathbf{cfo}$ and $\varphi_\mathbf{cpo}$ are the phase errors caused by cyclic shift diversity (CSD), sampling frequency offset (SFO), symbol timing offset (STO), carrier frequency offset (CFO) and carrier phase offset (CPO), respectively. $Z$ signifies the additive measurement noise. In the following, we will address these deep-rooted CSI phase issues in a divide-and-conquer manner.
\begin{figure*}[!t]
	\centering
	\includegraphics[width=1\linewidth]{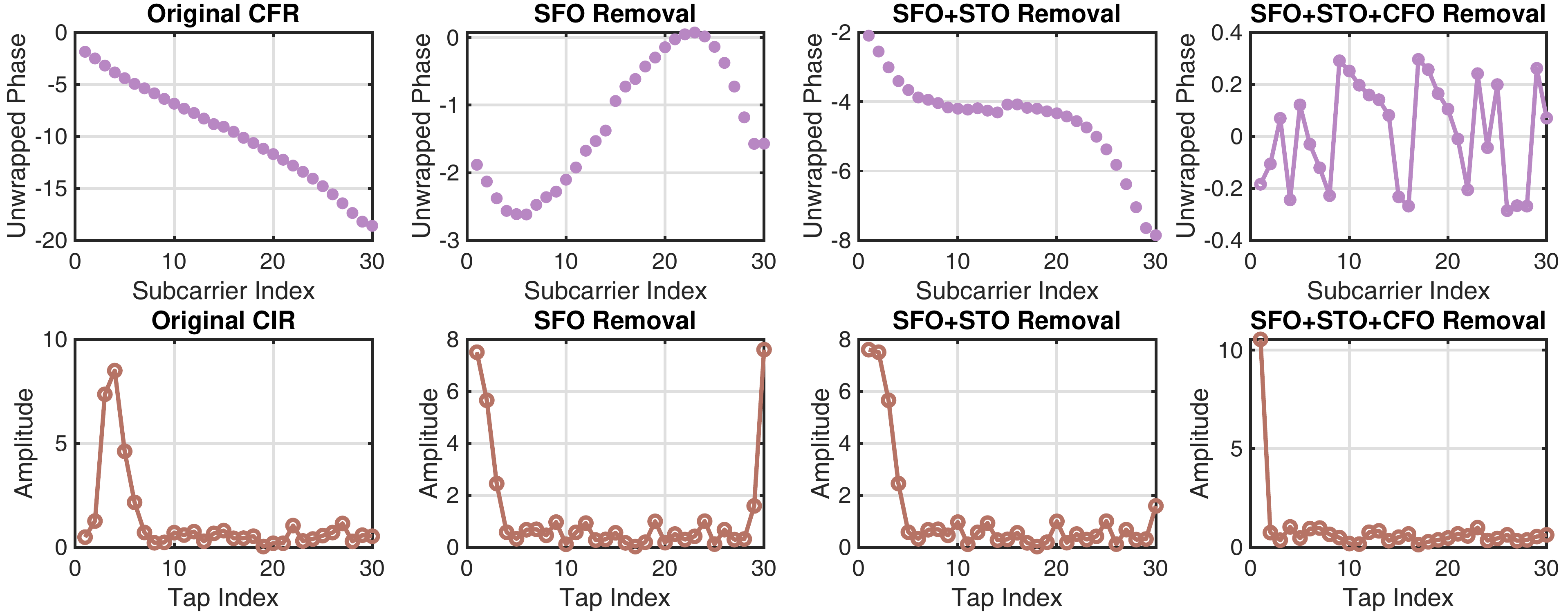}
	\caption{The CFR phase (Row 1) and CIR amplitude (Row 2) changes after SFO, STO and CFO removal.}
	\label{phasecalib}
\end{figure*}
\begin{enumerate}
\item[$\bullet$] CSD: As described in Section \ref{subsec:mimo}, CSD is operated by sending cyclically shifted OFDM symbols over different TX antennas so that unintended beamforming is avoided. But this incurs an additive phase shift for each TX antenna in CSI matrix which potentially degrades the localization performance. SignFi \cite{ma2018signfi} compensated the CSD errors by applying a multiple linear regression scheme. However, as a easier alternative suggested in \cite{tadayon2019decimeter}, CSD can always be removed by the receiver when direct mapping takes place, under which the SMM equals an unitary matrix. Hence, in our data acquisition process, we can configure the Intel 5300 shipset to make $N_{ss}=N_t$, thus yielding the CSD-free CSIs.

\item[$\bullet$] SFO: In OFDM transceiver system, SFO occurs when the receiver's ADC sampling rate differs from the transmitter's synthesization rate. Consequently, SFO manifests itself as an additive phase shift proportional to the subcarrier index, which gives rise to the first-order channel linearity (e.g. Fig. \ref{phasecalib}). We then resort to a simple linear regression method to remove the residual SFO. It can be mathematically expressed as follows.
\begin{equation}
\widehat{\xi}_\mathbf{sfo}= \argmin_{\rho} \sum_{k,n_t,n_r} (\varphi_{n_t,n_r}^k + 2\pi f_\delta k \rho + \omega)^2
\end{equation}
where $\rho$ and $\omega$ are curve fitting variables and $\varphi_{n_t,n_r}^k$ denotes the unwrapped CSI phase for one packet at the $k^{th}$ subcarrier from $n_t^{th}$ TX antenna to $n_r^{th}$ RX antenna, $n_t \in [1,N_t]$ and $n_r \in[1,N_r]$.

\item[$\bullet$] STO: In general, the receiver utilizes the auto/cross-correlator to capture and detect the presence of the OFDM symbol header, which starts with short training fields (STFs). However, the length limitation of these STFs brings great uncertainty to determine the symbol boundary. This results in the irreversible STO. Fortunately, given that any frequency domain phase shift due to STO leads to the same amount of circular rotation in time domain, STO can be embodied as peaks at the far end of power delay profile (PDP) owing to the CIR's cyclic-shifting property. On this basis, in order to estimate STO, we first derive the PDP from the CIR vector, i.e. $\{ h_k (n)\}_{1 \leq k \leq K}$ of the $n^{th}$ packet. The corresponding tap index $N_{sto}(n)$ of the shifted PDP peak due to STO can be identified as
\begin{equation}\label{equ:sto}
N_{sto}(n) = \argmax_{k} | h_k (n) |^2, 1 < k \leq K
\end{equation}
After applying (\ref{equ:sto}) for multiple packets, the most frequent value of $N_{sto}$ is then determined to finally shape the estimated STO as $\widehat{\varphi}_\mathbf{sto}=-2\pi k N_{sto}/ K$.

\item[$\bullet$] CFO/CPO: Due to the residual errors in receiver's phase locked loop (PLL), CFO emerges when the receiver's carrier frequency for down-conversion mismatches with the transmitted carrier frequency. Meanwhile,
since each time when the synthesizer restarts, a random initial phase will be generated by the receiver's voltage controlled oscillator and PLL cannot fully compensate for this phase difference, CPO is then experienced. According to \cite{tadayon2019decimeter}, after the PDP-based STO removal, the ToF estimation becomes naturally immune to CPO. Additionally, during our site survey, we only initiate the transceiver devices for once, which makes CPO negligible in our fingerprinting system. As CFO is also an accumulative error that has to be compensated by the receiver, we then employ a non-overlapping moving window with length $N_p$ for geometric averaging to further smooth out CFO. Specifically, we first obtain  $K$-dimensional $\bar{\mathbf{H}}$ by conducting element-wise multiplication for $N_p$ packets.
\begin{equation}
%\begin{align}
\bar{\mathbf{H}}=\mathbf{H}(1) \circ \cdots \mathbf{H}(n_p) \cdots \mathbf{H}(N_p), n_p \in [1,N_p]
%\end{align}
\end{equation}
where $\circ$ denotes the Hadamard product operator and $\mathbf{H}(n_p)$ is the $n_p^{th}$ CFR packet. The sanitized CFR can be then acquired by $\widehat{\mathbf{H}}=\{ (\bar{{H}}_k)^{\frac{1}{N_p}} \}_{1 \leq k \leq K}$.
\end{enumerate}

As illustrated in Fig. \ref{phasecalib}, the above adopted phase calibration techniques have effectively compensated CSI phase errors after the respective SFO, STO and CFO removal. 

%%%%%%%%%%%%%%%%%%%%%%%%%%%%%%%%%%%%%%%%%%%%%%%%%%%%%%%%%%%%%%%%%%%%%%%%%%%%%%%%%%%%%%%%%%%%%%%%%%%%%%%%%%%%
\subsection{Offline Radio Map Construction}
After the noise removal and phase sanitization, the pre-processed CSIs then proceed readily to establish a self-contained fingerprint database which involves both amplitude and phase information.

\subsubsection{\textbf{AR Entropy Estimation using CSI Amplitude}}

Recall that the entropy metric is deemed as a desired location fingerprint due to its structural simplicity as well as its statistical embodiment of rich location-specific information. In reality, it is intractable to directly derive Shannon entropy from real data \cite{bercher2000estimating}. The reason behind this dilemma is twofold: (i) Given that fact that the true PDF is normally unknown, entropy approximation is only reachable from the mere data samples. (ii) Conventional Shannon entropy calculation requires cumbersome numerical integration since a closed-form substitute does not exist.

To address the above challenges, in this paper, we propose to accurately estimate the entropy by leveraging AR modeling approach, whose core principle is to estimate the PDF-equivalent PSD of an unit variance AR process. This unit variance constraint is imported to meet the basic requirements of PDF (i.e. positive function that integrates to one). Specifically, we define a general notation of CSI amplitude $\beta$ from one subcarrier and the entire PDF-PSD relations is presented by
\begin{equation}
\label{equ:psd}
p(\beta)=S_{W}(\beta)=\frac{\sigma_{\epsilon}^{2}}{\left|1 + \sum_{i=1}^{p} a_{i} e^{-j 2 \pi i \beta}\right|^{2}},  \beta \in [-0.5 , 0.5]
\end{equation}
where $p(\beta)$ and $S_{W}(\beta)$ are the PDF and PSD of amplitude $\beta$, respectively. The set of $\{a_i\}_{1\leq i \leq p}$ are the AR coefficients of an order $p$ AR process and $\sigma_{\epsilon}^2$ is the model prediction error which is chosen so that $\int_{-0.5}^{0.5} S_W(\beta) d \beta =1$. It is notable that since the law is modeled as the spectrum restriction on the interval of $[-0.5,+0.5]$ \cite{kay1993fundamentals}, the amplitude data has to be first rescaled on this interval.

For solving Equation (\ref{equ:psd}), a proper model order $p$ should be first chosen since a low order leads to inadequate resolution and a high order incurs spurious peaks (excessive variance). We thereby adopt the exponentially embedded family (EEF) technique \cite{kay2005exponentially} to determine the AR model order due to its superior performance through our extensive experiments. Subsequently, both the AR coefficients and its corresponding model prediction error can be estimated by solving the well-known Yule-Walker equations using Levinson-Durbin recursion \cite{luan2019entropy}. Once the AR PDF is determined from (\ref{equ:psd}), the entropy computation can be then expressed by the following form:
\begin{equation}
\label{equ:arentropy}
\begin{split} 
\widehat{\phi}_{\beta}  =& -\int_{-0.5}^{0.5} \widehat{p}(\beta) \log \widehat{p}(\beta) d \beta \\
                        =& -\int_{-0.5}^{0.5} \widehat{S}_{W}(\beta) \log \widehat{S}_{W}(\beta) d \beta
\end{split}
\end{equation}
Besides, by applying Plancherel-Parseval formula to the right hand side of (\ref{equ:arentropy}), a feasible closed-form alternative without any numerical integration can be yielded as
\begin{equation}
\label{equ:lastentropy}
\widehat{\phi}_{\beta}=-\sum_{i=-\infty}^{\infty} R_{W}(i) Z_{W}^{*}(i)
\end{equation}
where $(\cdot)^{*}$ is the conjugate operator and $Z_{W}(i)$ denotes the $i^{th}$ component of the AR process's cepstrum, which can be obtained by applying the IFFT to $\log \widehat{S}_{W}(\beta)$. $R_{W}(i)$ represents the autocorrelation function of the amplitude data.

\begin{figure}[!t]
	\centering
	\includegraphics[width=0.7\linewidth]{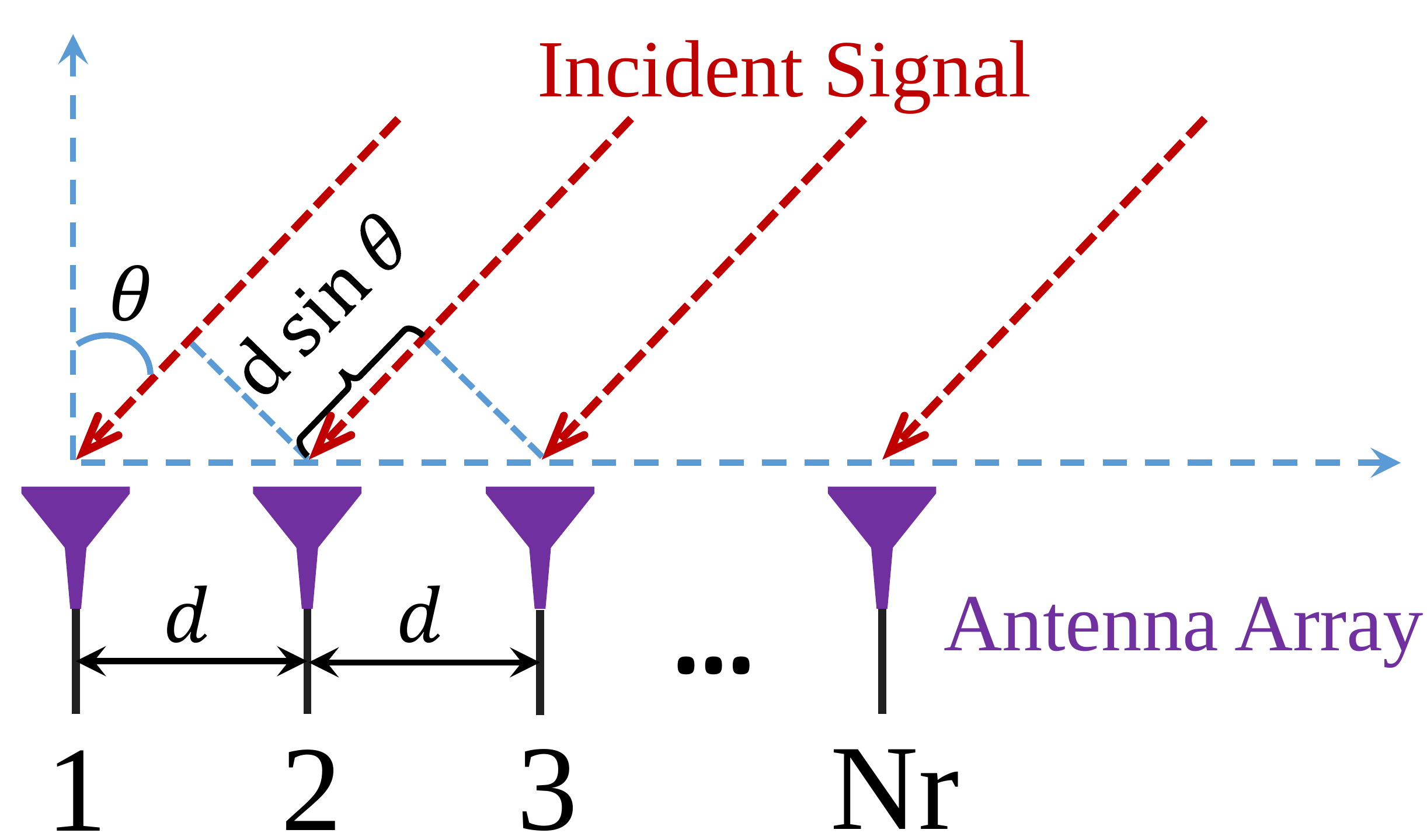}
	\caption{An incident signal arrives at an antenna array with an angle $\theta$}
	\label{array}
\end{figure}

In addition, as reported in \cite{luan2019entropy}, the AR entropy estimates from two endpoint subcarriers cause the most ambiguity for location distinction. We accordingly remove these subcarriers to further enhance the fingerprinting efficiency. Thus, when combining the total $R$-dimensional CSI measurements from all $N_t \cdot N_r$ antenna pairs, the final estimated AR entropy fingerprint in the offline phase can be restored as
\begin{equation}
\widehat{\bm{\Phi}}_{\mathbf{H}^s_m}=[\widehat{\phi}_{\mathbf{H}^s_m}^1, \dots, \widehat{\phi}_{\mathbf{H}^s_m}^r,\dots, \widehat{\phi}_{\mathbf{H}^s_m}^{R'}], r \in [1,{R'}]
\end{equation}
where $R'=R-2  N_t  N_r$ is the reduced fingerprint dimensionality in this case. Likewise, the online estimated AR entropy vector can be denoted by $\widehat{\bm{\Phi}}_{\mathbf{G}^s_{o}}=[\widehat{\phi}_{\mathbf{G}^s_{o}}^1, \dots, \widehat{\phi}_{\mathbf{G}^s_{o}}^r,\dots, \widehat{\phi}_{\mathbf{G}^s_{o}}^{R'}]$. Next, we will separately infer the AoA-based fingerprint by exploiting the CSI phase information. 

\subsubsection{\textbf{Enhanced AoA Estimation using CSI Phase}}
For considerable wireless location-aware applications, accurate AoA measurement is non-trivial on commodity devices. In view of the super-resolution advantage, the classical subspace-based MUSIC algorithm is of the greatest appeal. The basic idea of standard MUSIC algorithm is that incident signals from different bearings give rise to different phase changes on each antenna at the receiver \cite{schmidt1986multiple}. 

Assume that there are $L$ incoming signals $\gamma_1,\dots,\gamma_L$ arriving from directions $\theta_1,\dots,\theta_L$ at $N_r$ RX antennas of a linear array. The RX antennas are evenly-spaced with a distance $d$, which is about half of the signal's wavelength. As shown in Fig. \ref{array}, for the $l^{th}$ signal ($l \in [1,L]$), a phase difference of $-2\pi f d sin(\theta_l) / c$ is introduced at two adjacent antennas, where $f$ is the signal frequency and $c$ denotes the speed of light. For the whole antenna array, we can thereby define these phase shifts relative to the first antenna as the following steering vector.
\begin{equation}
\mathbf{\Psi}(\theta_l)=[1,e^{-j 2 \pi f d \sin(\theta_l)/c }, \dots, e^{-j 2 \pi (N_r-1) f d \sin(\theta_l)/c} ]^{\top}
\end{equation}
where $(\cdot)^{\top}$ is the transpose operator. Given all $L$ incident signal paths, the $N_r \times L$ steering matrix is then constructed by $\mathbf{Q}=[\mathbf{\Psi}(\theta_1),\mathbf{\Psi}(\theta_2),\dots,\mathbf{\Psi}(\theta_L)]$. Thus, the received signal $x$ at each RX antenna can be expressed as the superposition of all $L$ signal paths.
\begin{equation*}
[{x}_1,{x}_2,\dots,{x}_{N_r}]^{\top}= \mathbf{Q} [{\gamma}_1,{\gamma}_2,\dots, {\gamma}_{L}]^{\top} +\mathbf{W}
\end{equation*}
or
\begin{equation}\label{equ:music}
\mathbf{X}=\mathbf{Q} \bm{\Gamma} + \mathbf{W}
\end{equation}
where $\mathbf{W}$ is the noise vector.

Note that there is an inherent constraint when applying the conventional MUSIC algorithm to Eq. (\ref{equ:music}), which requires array antennas should outnumber the resolvable incident multipaths (i.e. $N_r>L$). However, in typical indoor environments, there are about 5-10 dominant multipath clusters \cite{czink2004number} while our commodity Intel 5300 NIC only supports up to $N_r=3$ antennas. This means it can merely capture 2 incident paths through MUSIC, thus largely limiting the AoA resolution and severely deteriorating the fidelity of the MUSIC outcome. To overcome this bottleneck, we leverage the fact that alongside AoA-related phase shifts across physical antennas, the incoming signals also invite phase differences across equispaced OFDM subcarriers due to ToF \cite{kotaru2015spotfi,vanderveen1997joint}. Therefore, we further extend the $N_r$-antenna physical array to a virtual sensor array with the size of $K \cdot N_r$, by which JADE-MUSIC algorithm can be readily employed to exploit CSI phase information in two dimensions. Specifically, the second steering vector which contains phase shifts relative to the first subcarrier can be defined as follows.
\begin{equation}
\mathbf{\Omega}(\tau_l)=[1,e^{-j 2 \pi  f_\delta \tau_l},\dots,e^{-j 2 \pi (K-1) f_\delta \tau_l}]^{\top}
\end{equation}
where $\tau_l$ is the time delay of the $l^{th}$ path and $f_\delta$ is the two adjacent subcarrier spacing. Accordingly, the combined AoA-ToF steering vector can be updated by
\begin{equation}
\mathbf{a}(\theta_l, \tau_l)= \mathbf{\Psi}(\theta_l) \otimes \mathbf{\Omega}(\tau_l)
\end{equation}
where $\otimes$ denotes the Kronecker product. After aggregating all $L$ signal multipaths, the corresponding $K N_r \times L$ steering matrix is thereby presented as
\begin{equation}
\mathbf{A}=[\mathbf{a}(\theta_1, \tau_1),\dots, \mathbf{a}(\theta_l, \tau_l), \dots, \mathbf{a}(\theta_L, \tau_L)]
\end{equation}
Hence, the received signals at RX antennas in Equation (\ref{equ:music}) can be rewritten by
\begin{equation}
\bar{\mathbf{X}}=\mathbf{A} \bm{\Gamma} + \bar{\mathbf{W}}
\end{equation}
Next, we then move to apply JADE-MUSIC by first deriving the covariance matrix $\mathbf{R}_{X}$ of the received signal, which is calculated as
\begin{equation}
\mathbf{R}_{X}=\mathbb{E} \{\bar{\mathbf{X}} \bar{\mathbf{X}}^{H}\}=\mathbf{A} \mathbf{R}_{S} \mathbf{A}^{H}+\sigma^{2}_W \mathbf{I}
\end{equation}
where $(\cdot)^H$ and $\mathbb{E} \{\cdot\}$ demotes the Hermitian transpose and expectation operator, respectively. $\mathbf{R}_{S}$ is the noise-free covariance matrix of the complex signal vector and $\sigma^{2}_W$ indicates the noise variance. Among $K\cdot N_r$ eigenvalues of $\mathbf{R}_{X}$, the smallest $(K N_r-L)$ eigenvalues represent the noise and the remaining $L$ eigenvalues correspond to $L$ incident signals. The eigenvectors corresponding to these smallest eigenvalues then form the noise subspace $\mathbf{E}_{N}$. Since the signal subspace and noise subspace are orthogonal, the spatial pseudo-spectrum function can be expressed as follows.
\begin{equation}
P(\theta, \tau)=\frac{\mathbf{a}^H (\theta,\tau) \mathbf{a}(\theta,\tau)}{\mathbf{a}^H (\theta,\tau) \mathbf{E}_{N} \mathbf{E}_{N}^{H} \mathbf{a}(\theta,\tau)}
\end{equation}
By searching on the 2-D angle and delay continua, the sharp peaks in $P(\theta, \tau)$ will occur at the bearings of incident signals with their corresponding time delays.

$a)$ \textit{Forward-Backward Spatial Smoothing}: In practice, subspace techniques like MUSIC also require the signal covariance matrix $\mathbf{R}_{S}$ has full rank. However, our stacked CSI measurements $\bar{\mathbf{X}}$ from all the subcarriers at all RX antennas is just a single column unit rank matrix. Due to the coherence of multiple signals, all subspace based methods suffer complete failure from the rank deficiency of $\mathbf{R}_{S}$. To address this issue, we propose to apply forward-backward spatial smoothing to mitigate the random noise and further improve the joint AoA-ToF estimation performance \cite{pillai1989forward}. 
\begin{figure}[!t]
	\centering
	\includegraphics[width=1\linewidth]{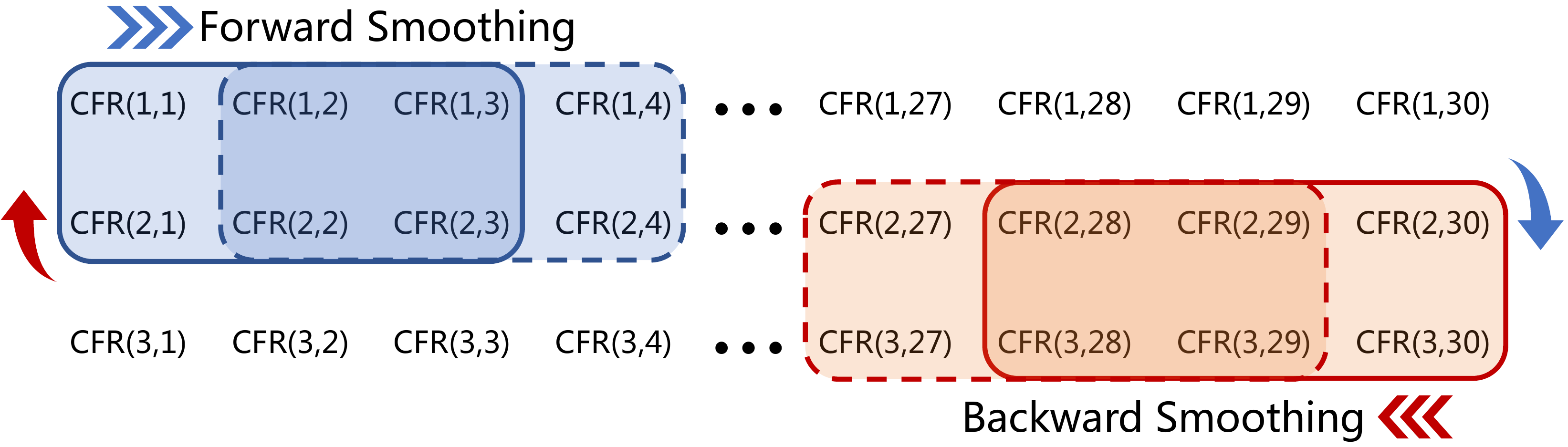}
	\caption{The mechanism of forward-backward smoothing for the case $K=30$ and $N_r=3$.}
	\label{smoothing_fb}
\end{figure}
%when consider subcarrier k=0, $\widehat{N}_s=N'_s-(K'-1)(N_r-N'_r+1)=(K-2K'+2)(N_r-N'_r+1)$
As shown in Fig. \ref{smoothing_fb}, after reshaping the single measurement vector to the $N_r \times K$ CSI matrix, we first partition the CSI matrix into uniformly overlapping subarrays with the size of $K' N'_r$, where $K'$ and $N'_r$ are the number of subcarriers and antennas in the subarray, respectively. To ensure measurable phase shifts across RX antennas, here $N'_r$ is fixed as 2 in our case. The total number of overlapping subarrays is then $T_K T_N$, where $T_K=K-K'+1$ and $T_N=N_r-N'_r+1$. In the sequel, a hardened spatially smoothed covariance matrix can be derived by averaging across those subarrays' covariance matrices with a forward direction (blue arrow). It is defined as follows.
\begin{equation}
\mathbf{R}_{f}=\dfrac{1}{T_K T_N} \sum_{i=1}^{T_K T_N} \mathbf{R}^i_{s}
\end{equation}
where $\mathbf{R}^i_{s}$ is the covariance matrix of the $i^{th}$ subarray. This covariance hardening processing achieves an improved rank, thus closer to the true source covariance matrix. Moreover, the invariant structure of CSI also enables a backward directional smoothing (red arrow) to further enhance the accuracy of MUSIC estimator. This averaged forward-backward covariance matrix can be expressed as
\begin{equation}
\mathbf{R}_{fb}=\dfrac{1}{2} (\mathbf{R}_{f}+\mathbf{J} \mathbf{R}^{*}_{f} \mathbf{J})=\dfrac{1}{2} (\mathbf{R}_{f} + \mathbf{R}_{b})
\end{equation}
where $\mathbf{J}$ is the $K' \times K'$ exchange matrix with only ones on its anti-diagonal and $\mathbf{R}_{b}$ denotes the backward covariance matrix.

\begin{figure}[!t]
\centering
\includegraphics[width=0.9\linewidth]{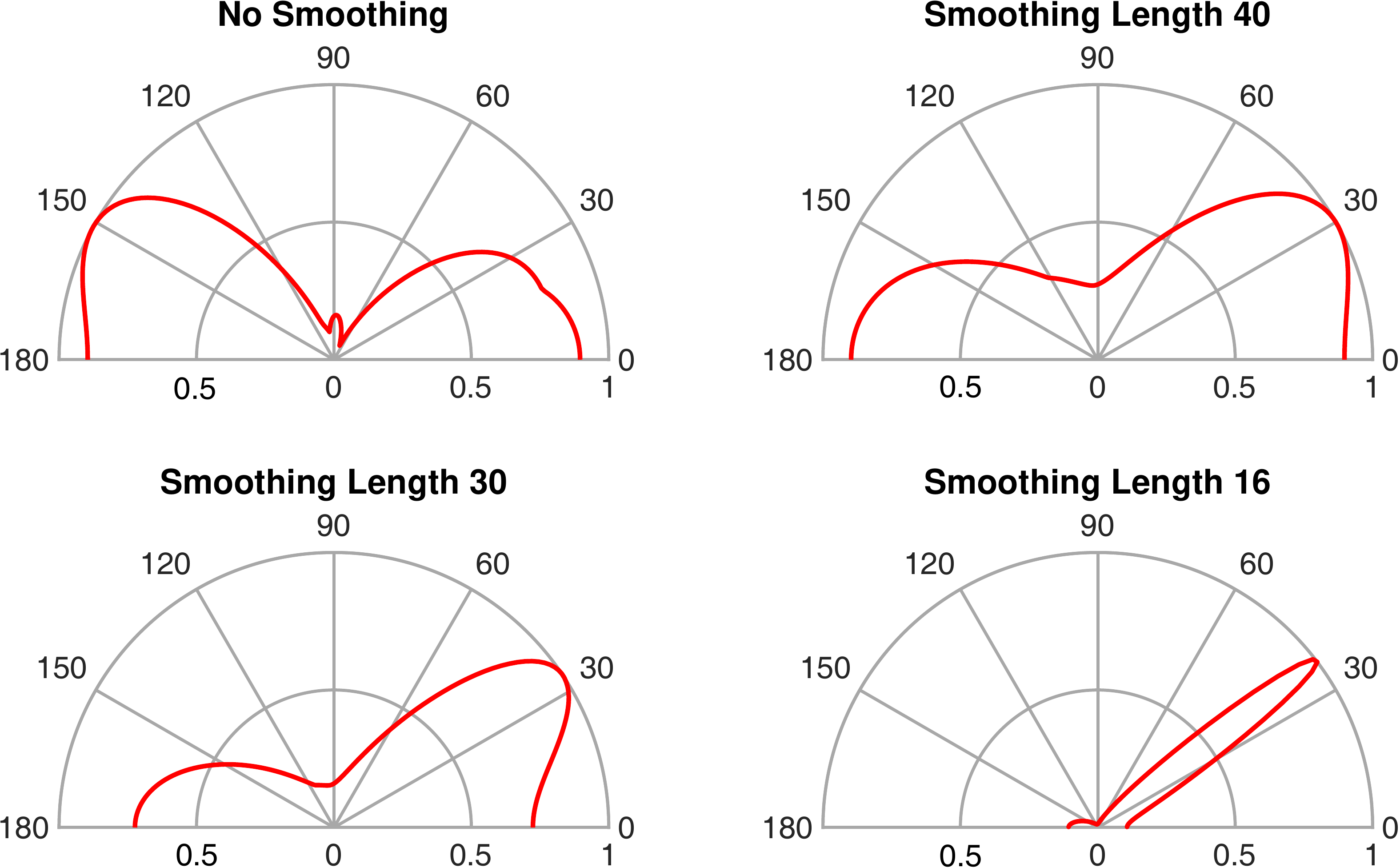}
\caption{Optimal smoothing length selection through AoA spectrum}
\label{smoothbest}
\end{figure}

$b)$ \textit{Optimal Smoothing Length Selection}: Note that SpotFi only treats smoothed CSI matrix with a fixed smoothing length of $K'N'_r=30$, which fails to dig deeper into the optimal selection of the smoothing length. As the smoothing length decreases, the noise level in estimated AoA spectrum gets lower, which helps to narrow the peak and improve the accuracy. But in the meantime, this also reduces the effective antenna sensors, which increases the risk of eliminating the peak from the direct path. To carefully cope with this trade-off problem, we perform a micro-benchmark which computes AoA spectra in a near LoS scenario (so the direct path bearing dominates) with different smoothing lengths. As observed in Fig. \ref{smoothbest}, the smoothing length of 16 ($K'=8$) shows a good compromise during our experiments and thus is chosen for the performance evaluation in Section \ref{sec:experiment}. It is also worth noting that since subcarrier index $k=0$ is null due to the large direct current (DC), in addition to making smoothing length larger than the number of multipaths indoors (say 10), we also need to ensure that no partitioned subarray contains $k=0$ subcarrier, which avoids $2 f_\delta$ error for AoA estimation.

\begin{figure}[!t]
	\centering
	\includegraphics[width=0.7\linewidth]{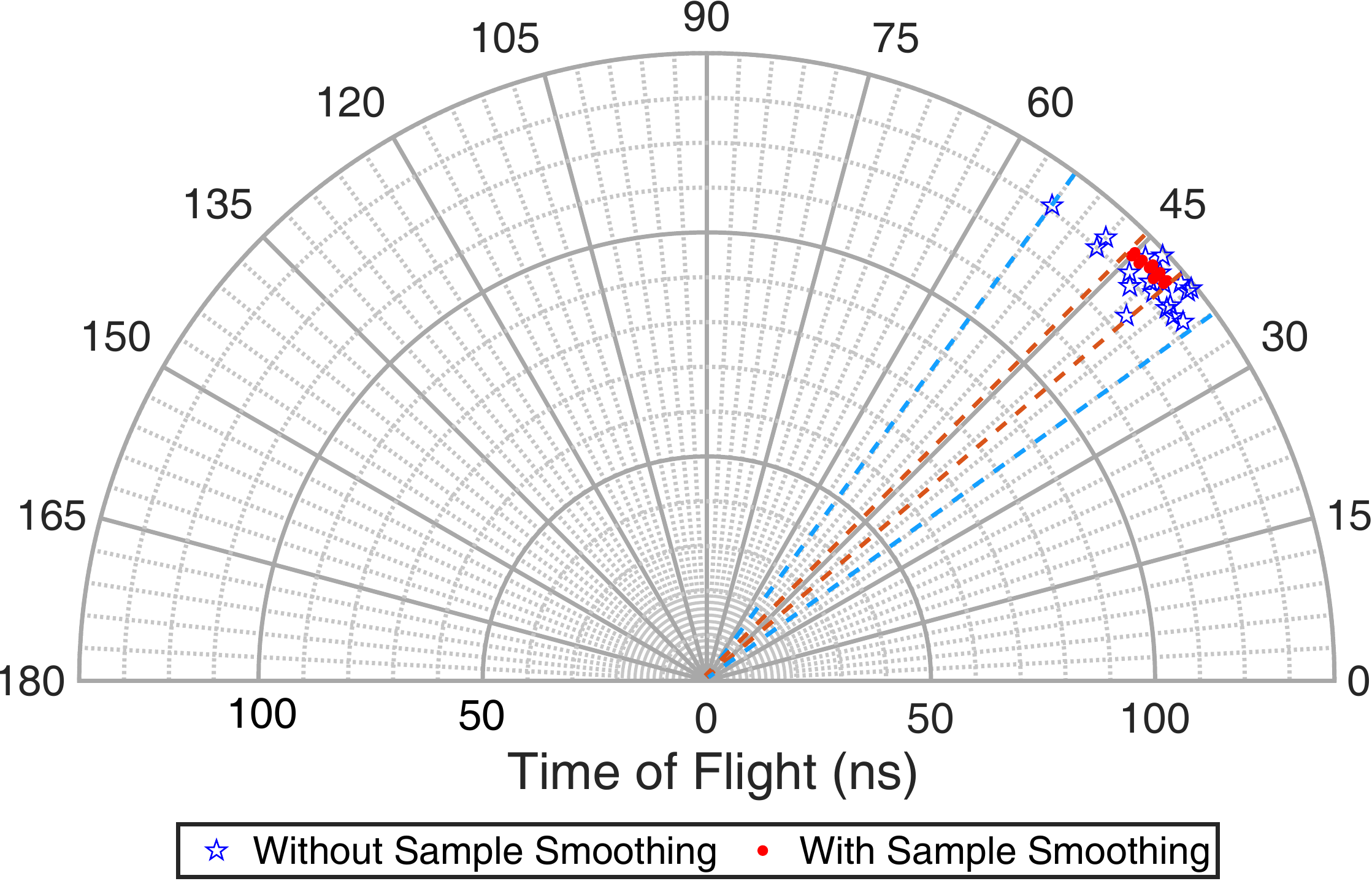}
	\caption{The comparison of AoA-ToF estimation with and without multi-packet smoothing.}
	\label{aoamultipacket}
\end{figure}

$c)$ \textit{Augmented Multi-Packet Smoothing}: Considering that we only perform the forward-backward smoothing in the frequency domain, to fully acquire the empirical covariance matrix, different packet snapshots are also needed to implement the time-domain averaging, which can be denoted by
\begin{equation}
\mathbf{R}_{mp}= \dfrac{1}{N_{mp}} \sum_{i=1}^{N_{mp}} \mathbf{R}_{fb}^i
\end{equation}
where $N_{mp}$ is the number of multiple CSI packets. We can observe in Fig. \ref{aoamultipacket} that the joint AoA and ToF estimation is further refined after the process of multi-packet sample smoothing.

As a result, along with AR entropy fingerprints, the final estimated AoA with the smallest ToF at the $m^{th}$ RP location from the $s^{th}$ AP is determined and stored as $(\widehat{\theta}_m^s, \widehat{\tau}^s_m)$ for the following online position estimation. Likewise, the online acquired fingerprint can be expressed by $(\widehat{\theta}_o^s, \widehat{\tau}^s_o)$.

%%%%%%%%%%%%%%%%%%%%%%%%%%%%%%%%%%%%%%%%%%%%%%%%%%%%%%%%%%%%%%%%%%%%%%%%%%%%%%%%%%%%%%%%%%%%%%%%%%%%%%%%%%%%%

\subsection{Online Location Estimation}  
In the online location estimation phase, the mobile target is required to be accurately mapped with the pre-defined fingerprint database. In order to quantify the similarity between the offline stored fingerprints and the online measured CSIs, we manage to independently adopt two simple distance metrics for the respective AR entropy and AoA fingerprints. For amplitude based AR entropy, the Manhattan distance (a.k.a. taxicab metric) \cite{krause1986taxicab} is employed to measure the gap between two vectors through the summation of the absolute differences of their corresponding components. Given the offline and online entropy fingerprints $\widehat{\bm{\Phi}}_{\mathbf{H}^s_m}$ and $\widehat{\bm{\Phi}}_{\mathbf{G}^s_{o}}$, the Manhattan distance between them is represented as
\begin{equation}
\mathscr{D}_{m}^{s}= \| \widehat{\bm{\Phi}}_{\mathbf{H}^s_m} - \widehat{\bm{\Phi}}_{\mathbf{G}^s_{o}} \| ^{}_{1} =\sum_{i=1}^{R'}\left| \widehat{\phi}_{\mathbf{H}^s_m}^i - \widehat{\phi}_{\mathbf{G}^s_{o}}^i \right|
\end{equation}
where $\|\cdot\|_1$ denotes the $\ell_1$ norm. Moreover, by following the chain rule for Shannon entropy \cite{cover2012elements}, it can be proved that a joint entropy difference for multiple independent variables is equal to the sum of all these variable's entropy differences. Under the $S$ independent AP assumption, we therefore have the Manhattan distance for all available APs as follows.
\begin{equation}
\mathscr{D}_{m}=\sum_{s=1}^{S} \mathscr{D}_{m}^{s}
\end{equation}
For the estimated 2-D AoA and ToF fingerprints, we naturally resort to the simple Euclidean distance to capture the discrepancy between the offline $(\widehat{\theta}_m^s, \widehat{\tau}^s_m)$ and online $(\widehat{\theta}_o^s, \widehat{\tau}^s_o)$ from all $S$ APs. It can be then defined as
\begin{equation}
\mathcal{D}_m= \sqrt{ \sum \nolimits_{s=1}^S (( \widehat{\theta}_m^s-\widehat{\theta}_o^s)^2 + (\widehat{\tau}_m^s-\widehat{\tau}_o^s)^2 ) }
\end{equation}

In general, both of the two metrics are fully capable of concisely reflecting the spatial proximity between the offline learned traits at the $m^{th}$ RP and the online measurements from an uncharted position. For the design of AngLoc, the remaining location estimation process consists of two main steps. First, by adopting the classical kNN theory, we can claim $M_c$ out of $M$ RP locations which signify $M_c$ smallest AR entropy differences among $\{\mathscr{D}_{m}\}_{1 \leq m \leq M}$. Then, a novel bivariate kernel regression scheme is further proposed to infer the final target's location by exploiting the distance based kernel function and the selected set of $M_c$ reference points. The estimated location $\widehat{\bm{\ell}}_{o}$ is expressed by
\begin{equation}
\widehat{\bm{\ell}}_{o}=\frac{\sum_{m_c=1}^{M_c} \mathcal{K}_{m_c} \bm{\ell}_{m_c}}{\sum_{m_c=1}^{M_c} \mathcal{K}_{m_c}}
\end{equation}
where $m_c \in [1,M_c]$ and $\mathcal{K}_{m_c}$ denotes the probability kernel of the $m_c^{th}$ RP location which is obtained by exponentiating and weighting its corresponding entropy and AoA based distances. It can be mathematically presented as follows:
\begin{equation}
\mathcal{K}_{m_c} = w_{e} \exp (-\rho_{e} \mathscr{D}_{m_c}) + w_{a} \exp (-\rho_{a} \mathcal{D}_{m_c})
\end{equation}
Here $w_e$ and $w_a$ are the weighting factors for the respective AR entropy and AoA based kernel function and $w_{a}+w_{e}=1$. $\rho_e$ and $\rho_a$ are their corresponding kernel coefficients which are chosen to optimally minimize the fingerprinting error by leave-one-out cross-validation in the offline phase. It is noteworthy that this bivariate kernel $\mathcal{K}_{m_c}$ equals to one if the given two fingerprints are identical and decays to zero as the dissimilarity of two fingerprints increases. Simply put, this bivariate kernel provides a flexible way to naturally harness the CSI data and therefore makes full use of our probabilistic AR entropy and AoA information, thus leading to an improved localization performance.

The performance of our AngLoc fingerprinting system will be evaluated in the following section.
\begin{figure*}[!t]
	\begin{minipage}[t]{1\linewidth}
		\centering
		\subfloat[Testbed $\mypound 1$: CNAM lab scenario]{\includegraphics[width=0.25\linewidth]{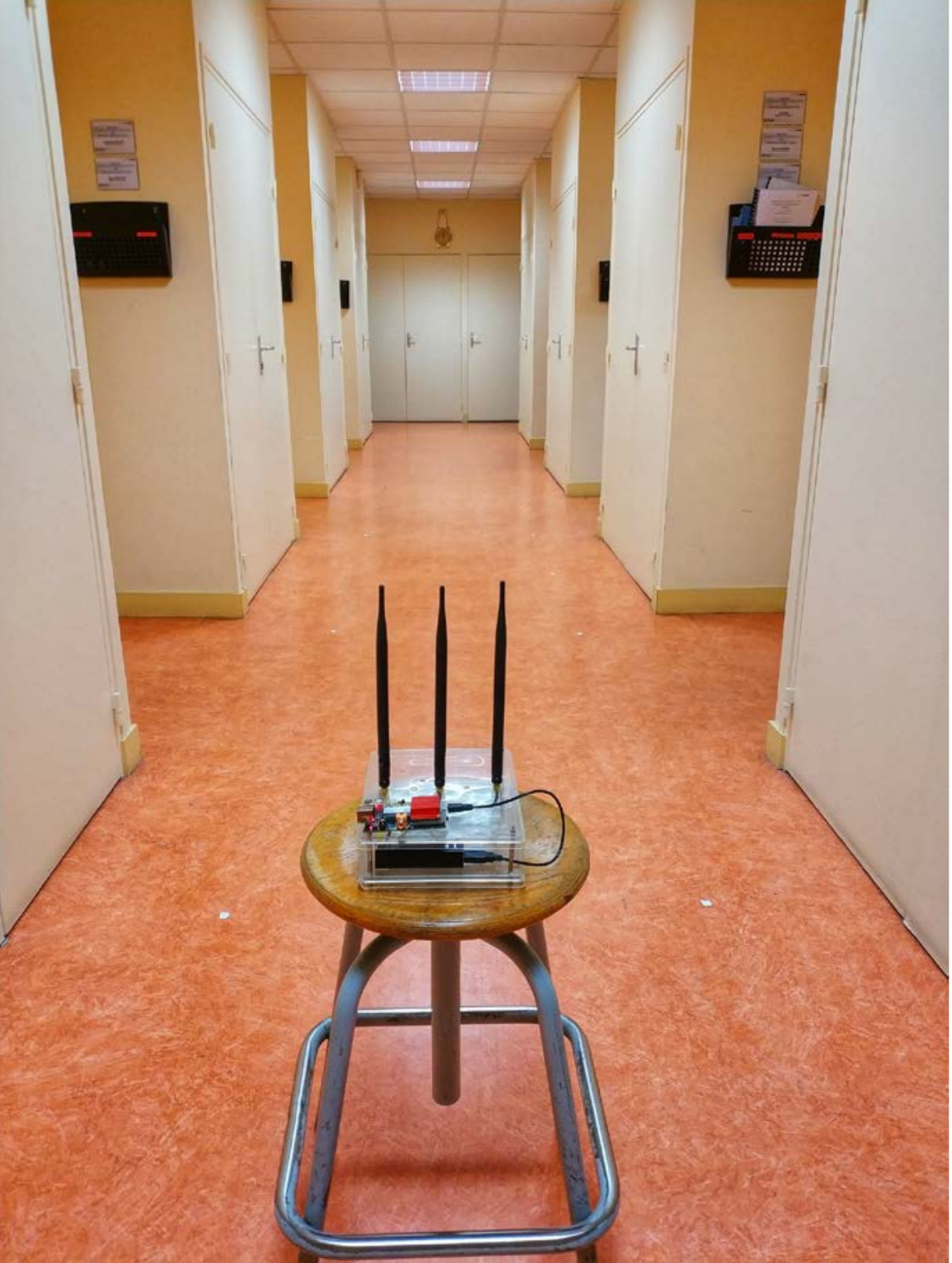}\label{lab}}
		%\hfill 
		\hspace{0.2cm}
		\begin{minipage}[b]{0.25\linewidth}
			\subfloat[TX: Laptop with Intel 5300 NIC]{\includegraphics[width=1\linewidth, height=1.02in]{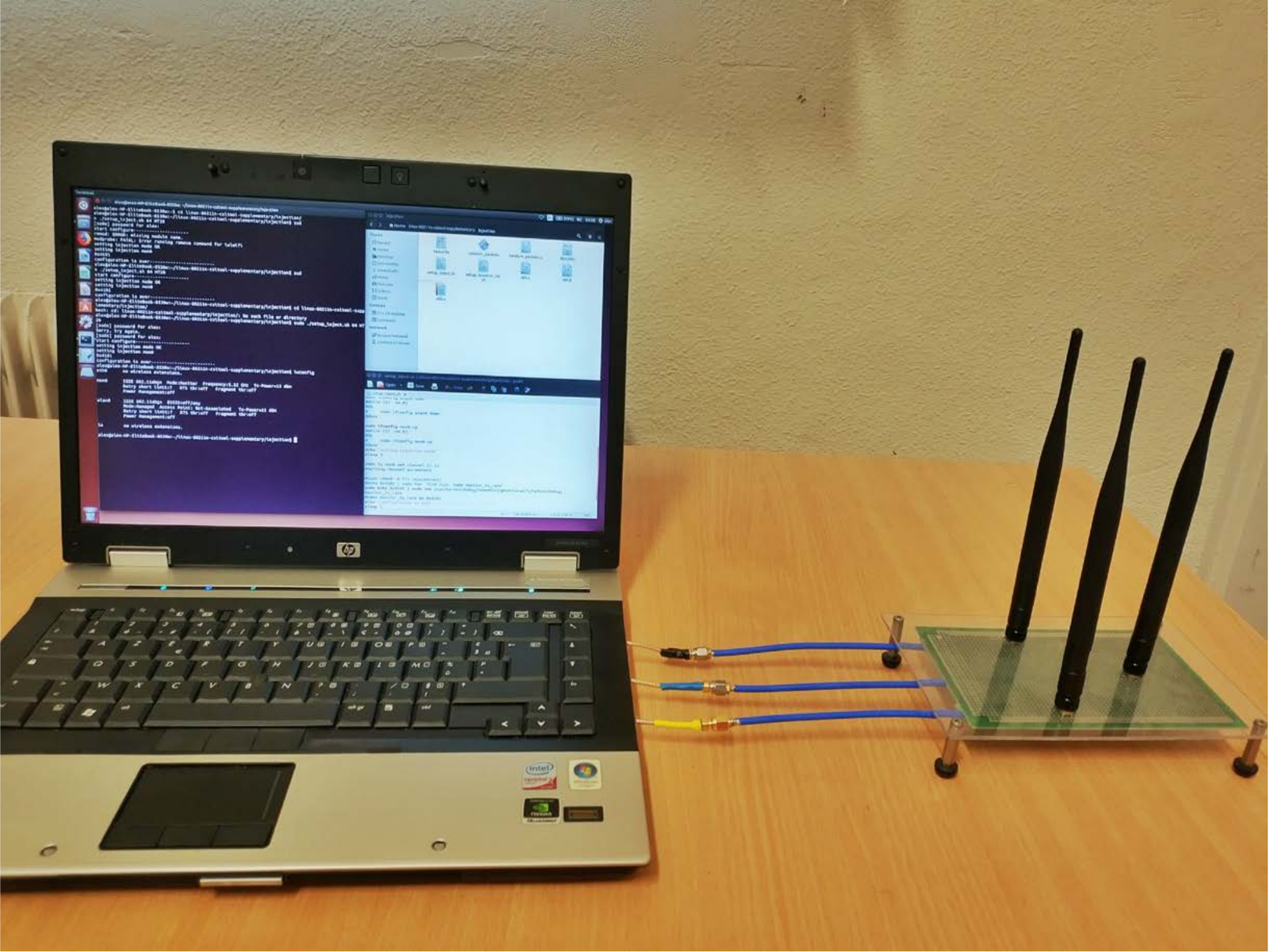}\label{pc}}
			\\%\vspace{0cm}
			\subfloat[RX: HummingBoard Pro]{\includegraphics[width=1\linewidth, height=1.02in]{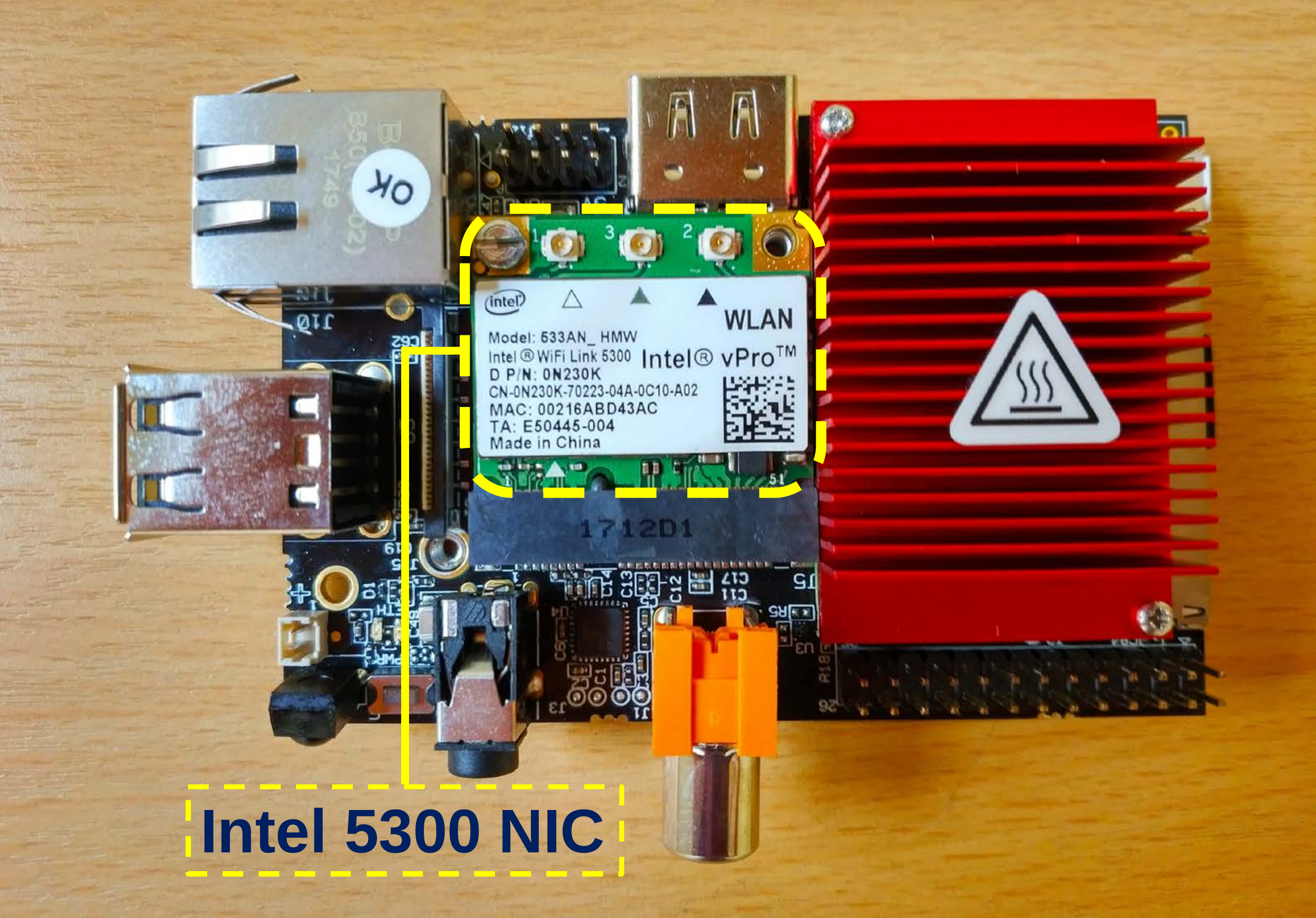}\label{hmb}}
		\end{minipage}
		%\hfill 
		\hspace{0.1cm}
		\subfloat[Testbed $\mypound 2$: Classroom scenario]{\includegraphics[width=0.25\linewidth]{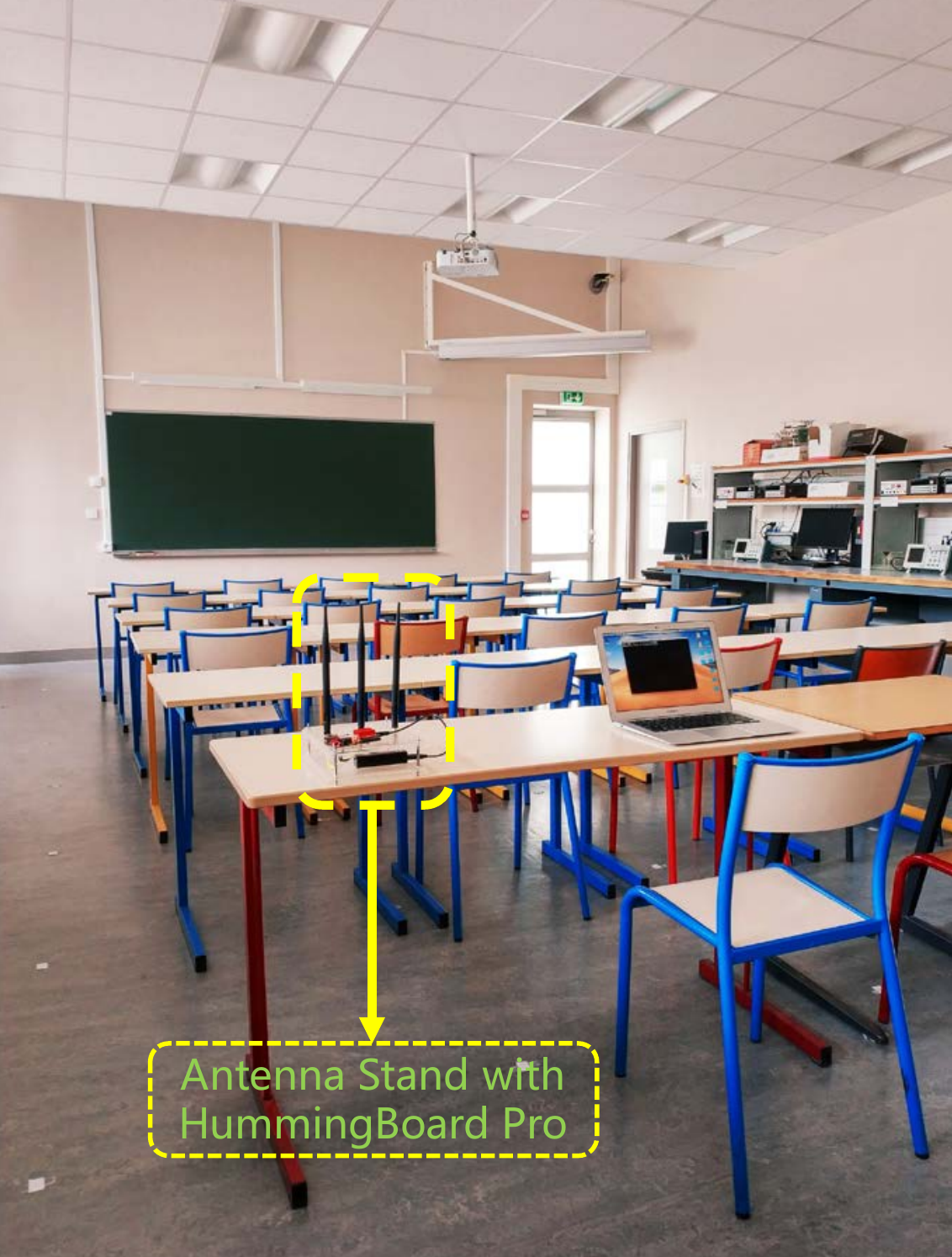}\label{classroom}}
		\caption{Two testbeds both with dedicated laptop as signal Transmitter (TX) and HummingBoard Pro as Receiver (RX): (a) CNAM laboratory scenario (Testbed $\mypound 1$); (b) TX: Laptop with Intel 5300 NIC; (c) RX: HummingBoard Pro; (d) CNAM classroom scenario (Testbed $\mypound 2$).}
	\end{minipage}\\	
	\begin{minipage}[t]{0.5\linewidth} 
		\centering
		\includegraphics[width=0.9\linewidth, height=2.2in]{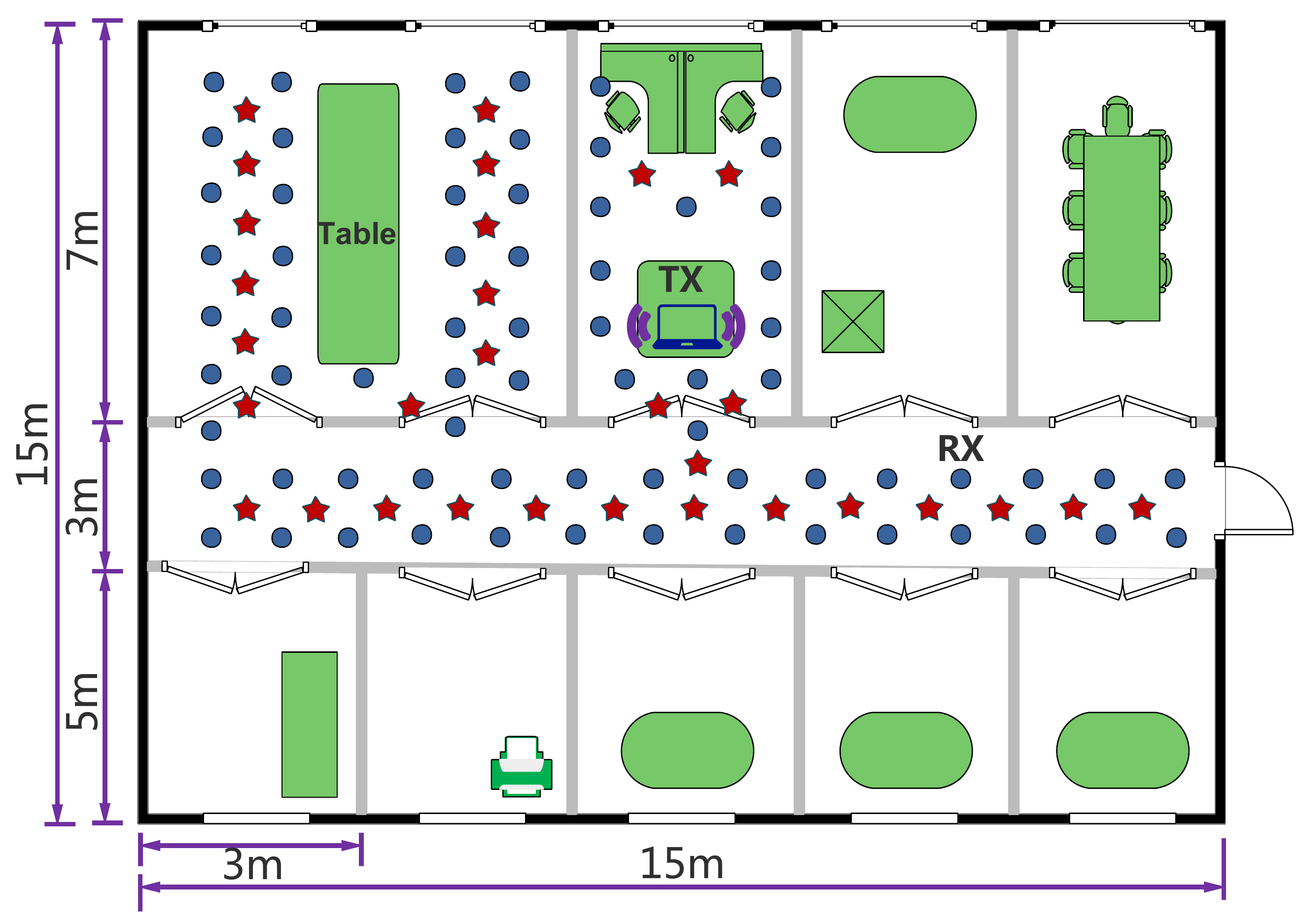}
		\caption{Floor plan of the laboratory (Testbed $\mypound 1$)} \label{fplan_lab}
	\end{minipage}
	\begin{minipage}[t]{0.5\linewidth} 
		\centering
		\includegraphics[width=0.9\linewidth, height=2.4in]{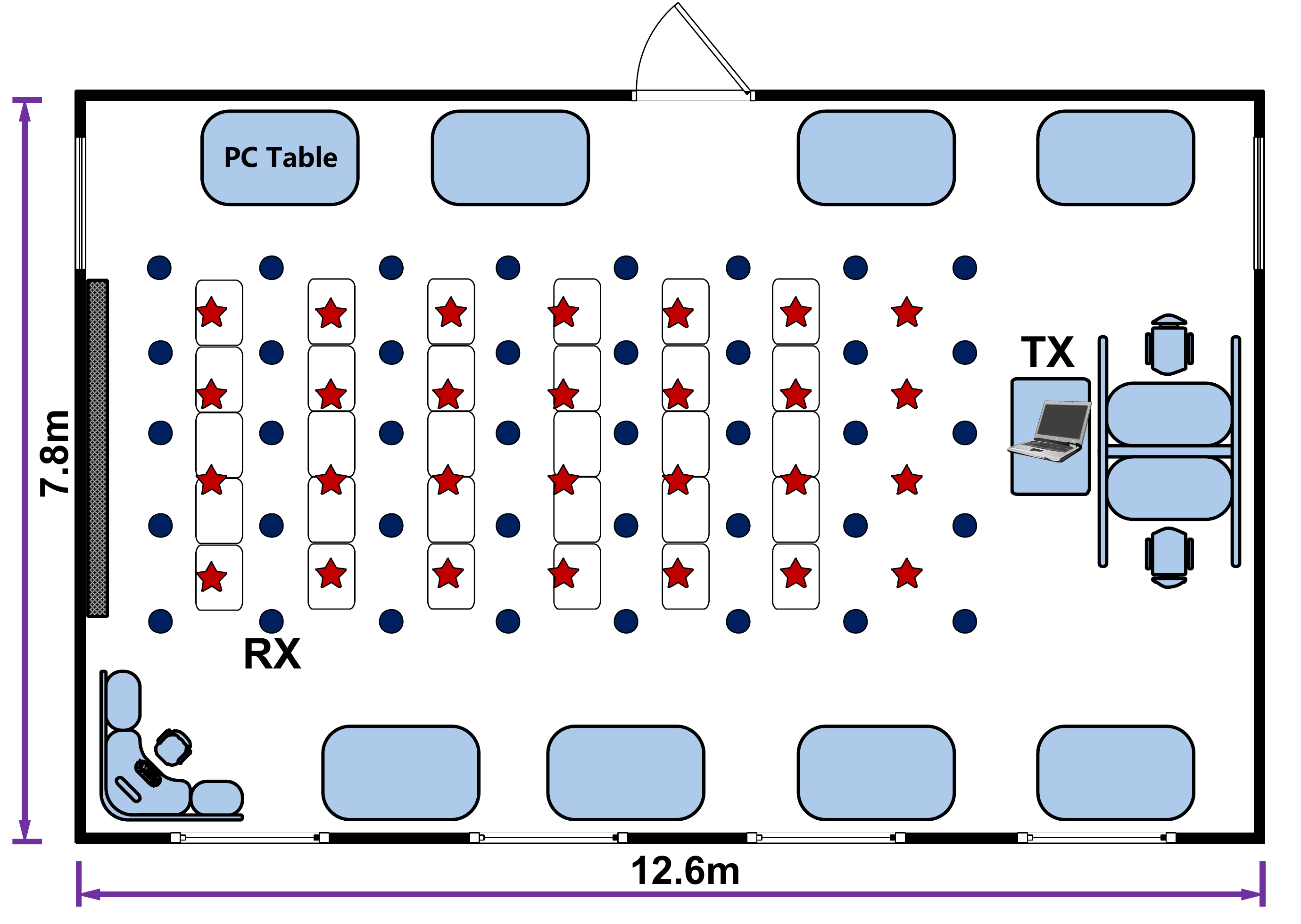}
		\caption{Floor plan of the classroom (Testbed $\mypound 2$)} \label{fplan_cr}
	\end{minipage}
\end{figure*}

%%%%%%%%%%%%%%%%%%%%%%%%%%%%%%%%%%%%%%%%%%%%%%%%%%%%%%%%%%%%%%%%%%%%%%%%%%%%%%%%%%%%%%%%%%%%%%%%%
\section{Performance Evaluation}\label{sec:experiment}
In this section, we carry out the experimental evaluation of our proposed localization system. We will begin with the experimental setup introduction and the detailed results of localization performance will be discussed in the sequel.

\subsection{Experimental Setup}
\subsubsection{Test Environments}
To evaluate the performance of our AngLoc system, the entire experiments are implemented at two different indoor testbeds in CNAM. As exhibited in Fig. \ref{lab}, the first testbed is a $15 m \times 15 m$ laboratory office in a multistorey building, which is comprised of a main corridor alongside several office and meeting rooms. Many desks, chairs, computers and shelves are furnished inside to form a complex indoor radio propagation environment. The second testbed in Fig. \ref{classroom} is an ample classroom scenario with an area of around 100 $m^2$. It lays out less obstacles within the fingerprinting area which presents a relative LoS scenario. It can then serve as a supplementary contrast with testbed $\mypound 1$.

\subsubsection{Hardware Descriptions}
For both testbeds, all the real experiments are conducted on the commodity-ready off-the-shelf Wi-Fi devices with a modified firmware \cite{Halperin_csitool}. To be specific, as shown in Fig. \ref{pc} and Fig. \ref{hmb}, by tuning into the IEEE 802.11n monitor mode with 5 GHz band, we deploy an HP Elitebook 8530w laptop as the signal transmitter and an Hummingboard Pro (HMB) device as the mobile receiver, both of which are equipped with Intel Wi-Fi Link (IWL) 5300 NIC and run 64-bit Ubuntu 14.04 OS and Debian 8.0 OS, respectively. In addition, as for the antenna settings, each wireless NIC-compatible device is capable of installing up to three omni-directional antennas so that the $3 \times 3$ MIMO configuration can be supported.

\subsubsection{Data Acquisition}
As aforementioned in the beginning, we implement the CSI data collections in both laboratory and classroom environments. Fig. \ref{fplan_lab} and Fig. \ref{fplan_cr} display the detailed floor plans and experimental layouts for our laboratory and classroom testbeds, respectively. First of all, for both testbeds, the laptop serves as signal transmitter whose placement is fixed on the table and known a priori. Under packet injection mode, it is designated to intermittently send at the rate of 100 packets per second using only one transmitting antenna. It is notable that such antenna setting means to meet the requirement of direct spatial mapping, which can yield CSD-free CSI data. Meanwhile, the localization accuracy can be also guaranteed with the lowest computational cost. For the two experimental layouts, the blue dots shown in Fig. \ref{fplan_lab} and Fig. \ref{fplan_cr} denote the 70(40) training RP locations with one meter spacing and the 30(28) testing positions are marked as red stars. During the offline training phase, roughly 5000 CSI packets are collected and stored by the lightweight HMB at each reference point to build up the raw CSI radio map. In the online phase, we proceed to move this HMB receiver among all the testing locations to acquire the same size of CSI packets for the localization purpose. Moreover, every receiver end is operated at the same height, constructing a simple 2-D platform for the precise indoor position estimation.

\subsubsection{Benchmarks and Performance Metrics}
In this section, we establish the whole benchmark program for the performance evaluation of our AngLoc system, which is compared with aforementioned systems like Horus \cite{youssef2005horus}, FIFS \cite{xiao2012fifs} and PinLoc \cite{sen2012you}. We also compare it with our previously proposed EntLoc system \cite{luan2019entropy}, which only exploits the CSI amplitude based entropy metric for indoor fingerprint localization. Besides, considering that the original PinLoc system conduct the war-driving procedure in a set of predefined $1 m \times 1 m$ grids, known as spots, in order to provide a fair comparison, we modify PinLoc to use the same training set that we use in the proposed AngLoc system. Particularly, for AoA accuracy evaluation, we take SpotFi \cite{kotaru2015spotfi} as the comparative rival due to its representativeness among recent AoA based IPSs.

As for the performance metrics, we define the localization error as Euclidean distance between the estimated location and the mobile user's actual position, which is presented as $\| \widehat{\bm{\ell}}_{o}-\bm{\ell}_{o}\|=\sqrt{(\widehat{x}_o-x_o)^2+(\widehat{y}_o-y_o)^2}$. When there are $N_{a}$ testing locations, we evaluate the localization performance by using the Mean Error (ME) metric which can be calculated as
\begin{equation}
ME=\dfrac{1}{N_a} \sum_{i=1}^{N_a} \sqrt{(\widehat{x}_i-x_i)^2+(\widehat{y}_i-y_i)^2}
\end{equation}
where $(x_i,y_i)$ and $(\widehat{x}_i,\widehat{y}_i)$ are the actual and estimated coordinates at the $i^{th}$ testing location, respectively.

\begin{figure}[!t]
	\centering
	\includegraphics[width=1\linewidth]{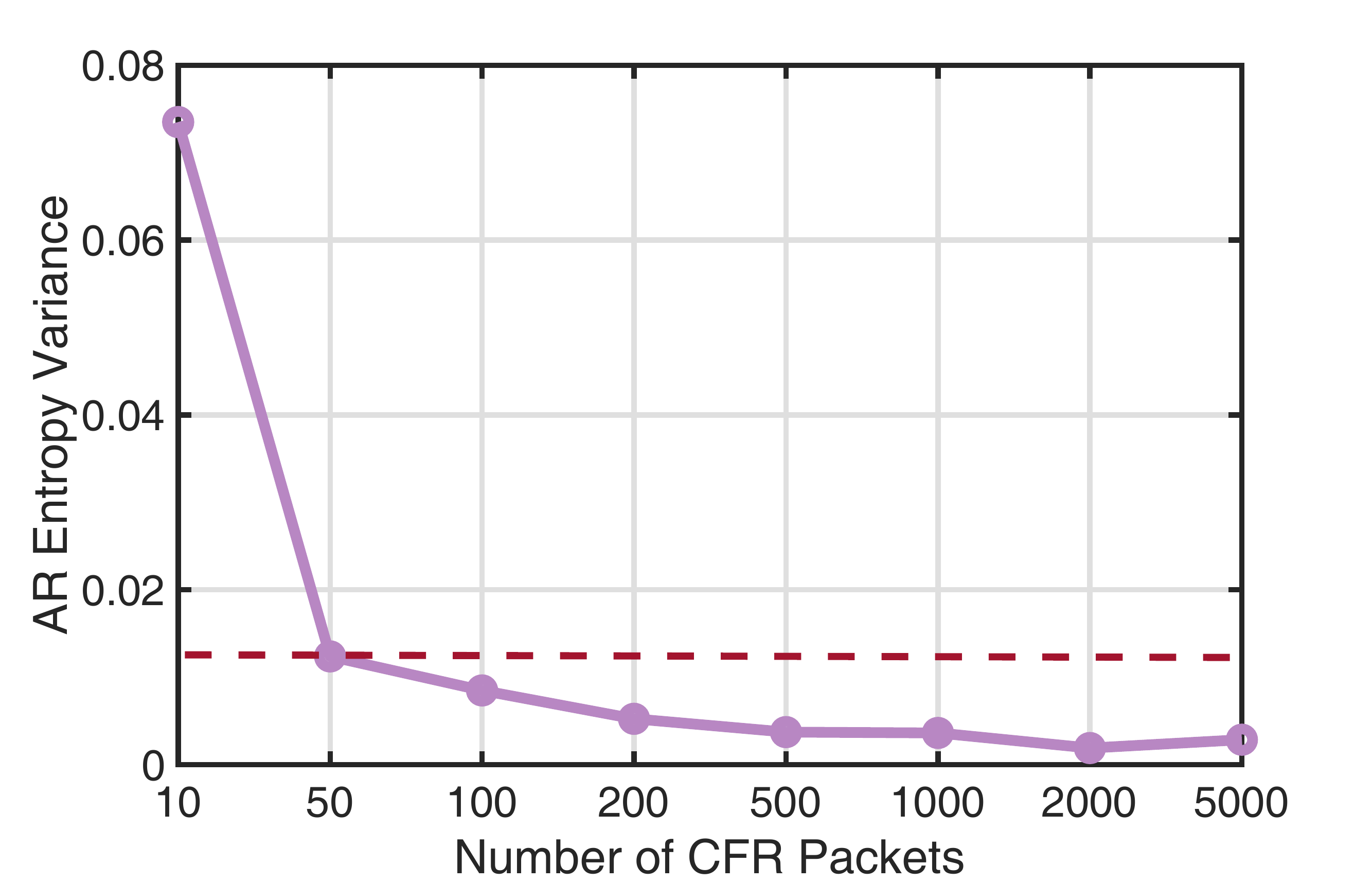}
	\caption{AR entropy variance changes with different CSI packet numbers.}
	\label{packnum_ent}
\end{figure}

\subsection{Experimental Results}
%\subsection{Effect of Different System Parameters}
Prior to exhibiting the localization results of our proposed system, we first reveal the effects of different system parameters which play a defining role for our AngLoc's performance. In addition, some other experimental factors will also be evaluated at the end of this section.

\subsubsection{Impact of Packet Number Selection for Entropy Estimation}
Since the AR entropy estimation process requires sufficient CSI samples, larger number of samples can provide more accurate entropy estimation but the computational cost ramps up. How to determine the CSI packet number for entropy calculation becomes a trade-off problem which needs to be balanced in our localization system. Accordingly, we put forward a variance based scheme to optimally select the number of CSI packets for AR entropy estimation. The motive lies in the fact that if the entropy variance is small enough, which is fairly stable to guarantee a good accuracy, there is no need to import more CSI samples to increase computational burden. Concretely, by changing the packet number ranging from 10 to 5000, we can observe in Fig. \ref{packnum_ent} that using 50 CSI packets can already provide stable enough AR entropy estimates, thus leading to the robust fingerprinting performance. So we select and fix this packet number for all entropy estimations in our indoor location implementations.

\begin{figure}[!t]
	\centering
	\includegraphics[width=1\linewidth]{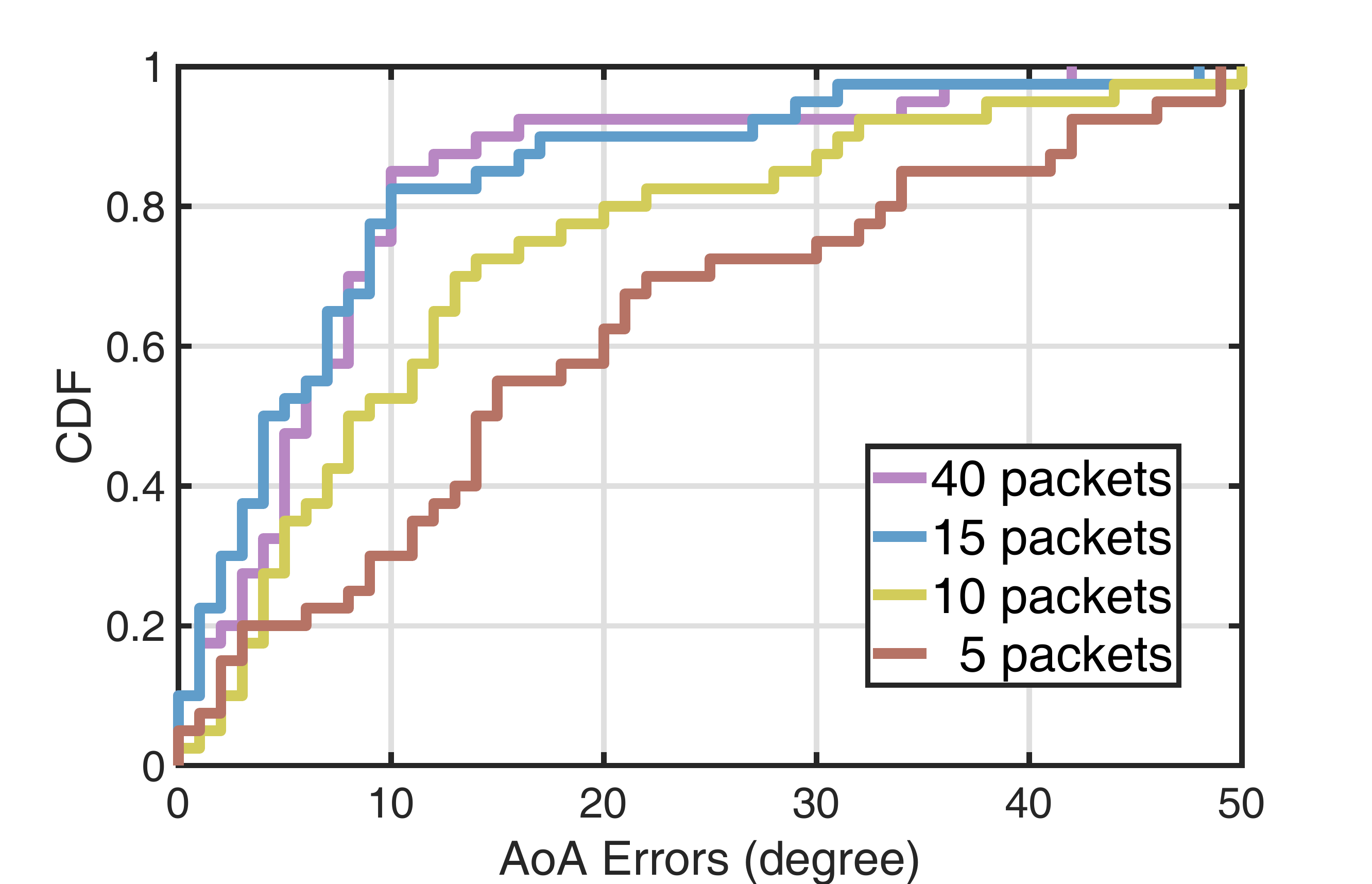}
	\caption{AoA estimation errors by using different number of CSI packets.}
	\label{packnum_aoa}
\end{figure}

\subsubsection{Impact of Packet Number Selection for AoA Estimation}
Due to the spatial-temporal diversity of CSI measurements, one CSI packet is able to derive the AoA estimate at one time. Likewise, the excess usage of CSI packets increases the unnecessary computational complexity while fewer number of CSIs risks generating more error-prone AoA estimations. On this basis, we devise an accuracy based packet number selection scheme to efficiently conduct AoA calculation. Considering that the multi-packet smoothing is required in our AngLoc system, we begin the testing packet number from 5 packets and extend it to 10, 15 and 40 packets, respectively. In addition, since the LoS-friendly classroom is more convenient and can provide a clear ground truth (direct path) to compare the AoA estimation errors. It is thus chosen as the experimental environment in this part. As shown in Fig. \ref{packnum_aoa}, we can observe that even with 15 packets, our AngLoc system works well and accurately identifies the true AoA with a mean error of 5 degrees, which shares the similar performance with 40 packets. The underlying explanation lies in the fact that once we determine the first arrival path through the smallest ToF, more CSI packets will not bring further improvement with regard to the AoA estimation accuracy. %Similar phenomenon can also be spotted in SpotFi paper.

\subsubsection{Impact of Kernel Regression Parameters}
Recall that in the online location estimation phase, we first find $M_c$ closest RP locations in accordance with the amplitude's AR entropy. Then, a weighted bivariate kernel regression scheme is proposed to accurately calculate the target's location by exploiting both entropy and AoA informations. As a result, a proper selection of the relevant kernel regression parameters in the offline phase is of great importance in the final localization outcome. As listed in Table \ref{paratable}, by leveraging leave-one-out cross-validation, we optimally choose the $M_c$, weighting factors $w_e$, $w_a$ and kernel coefficients $\rho_e$, $\rho_a$ for both testbeds. It is interesting to observe that in the larger and more NLoS laboratory scenario, the AoA-driven RP refining scheme outweighs the AR entropy factor (i.e. $w_a > w_e$), which indicates the fact that the entropy metric tends to bring more ambiguities in more complex environment. On the contrary, the more LoS classroom testbed renders the AR entropy competent enough to differentiate locations since the channel property in such case appears to be more stable.

\begin{figure}[!t]
	\centering
	\subfloat[Testbed $\mypound 1$]{\includegraphics[width=1\linewidth]{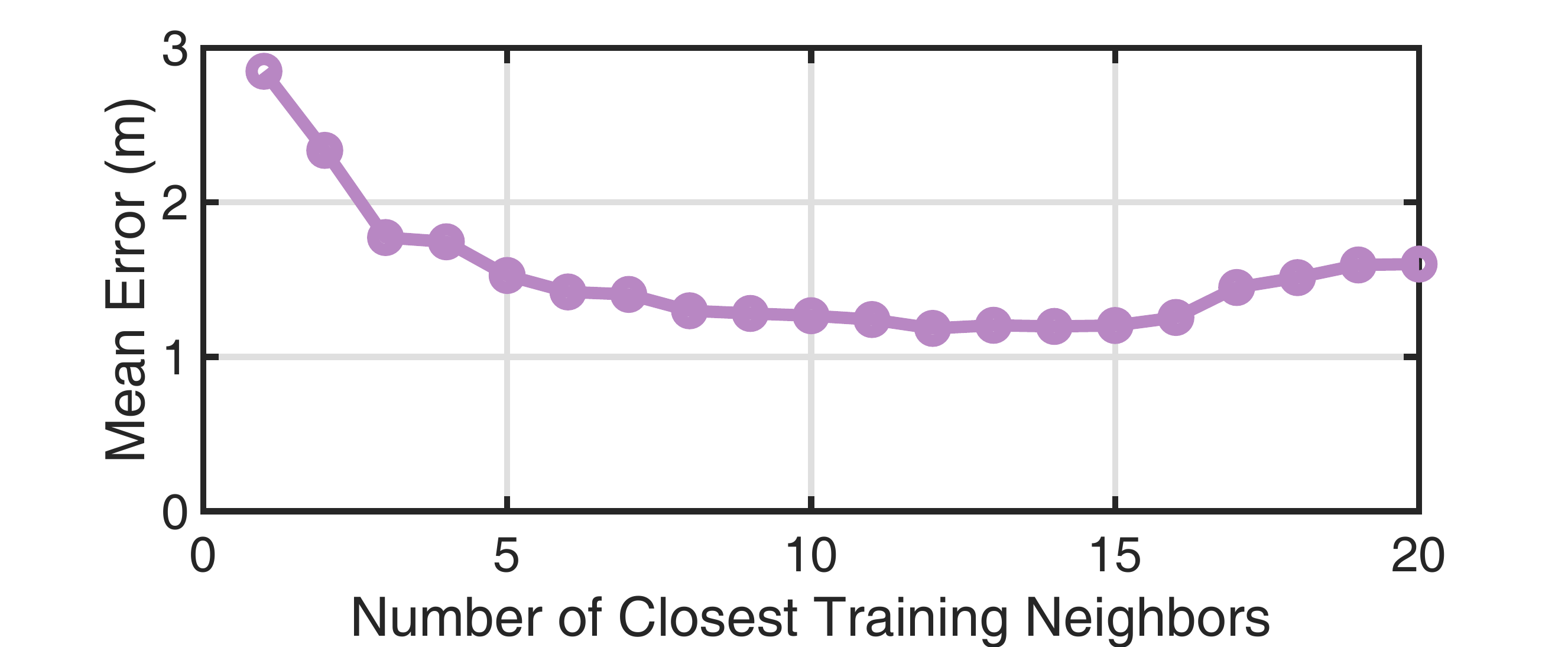}
		\label{bestk1}}
	\hfil 
	\subfloat[Testbed $\mypound 2$]{\includegraphics[width=1\linewidth]{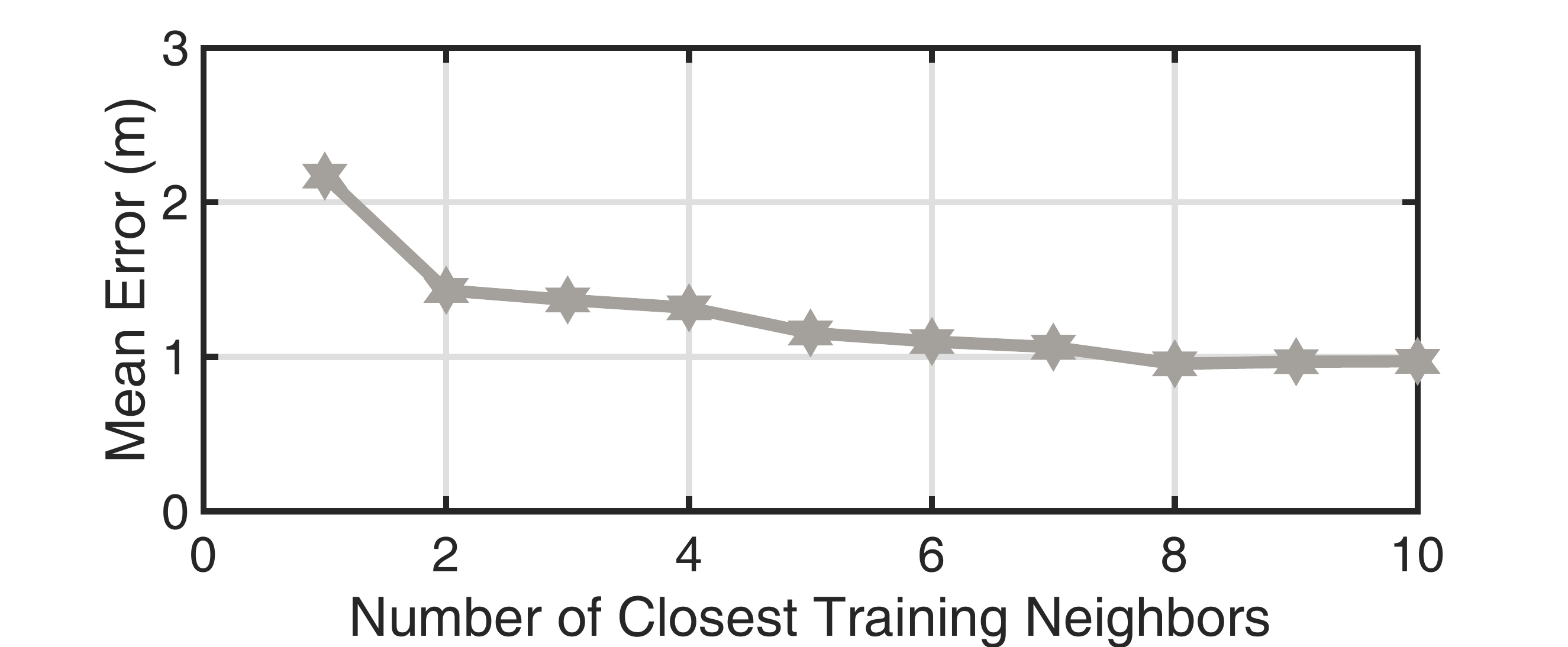}
		\label{bestk2}} 
	\caption{Optimal neighbor number selection for both (a) Testbed $\mypound 1$ and (b) Testbed $\mypound 2$.}
\end{figure}

\begin{table}[t!]
	\centering
	\caption{\textbf{Summary of parameters for both testbeds}}
	\label{paratable}
	\bgroup
	\def\arraystretch{2}%  1 is the default, change whatever you need; change table gap (Luan)
	\begin{tabular}{c|p{0.6cm}<{\centering}|c|c|c|c}
		\hline
		Parameters              &  $M_c$  &  $w_{a}$ &  $w_{e}$ &  $\rho_{a}$ & $\rho_{e}$  \\
		\hline \hline
		Testbed $\mypound 1$    &  12   &  0.57      &   0.43     &    0.14       &    0.23       \\
		\hline
		Testbed $\mypound 2$    &   8   &  0.38      &   0.62     &    0.33       &    0.17       \\
		\hline
	\end{tabular}
	\egroup
\end{table}

Furthermore, we also lay out the training results for choosing parameter $M_c$ in Fig. \ref{bestk1} and Fig. \ref{bestk2} from both laboratory and classroom environments. Given a respective range of $[1, 20]$ and $[1,10]$, we can identify the optimal selection of $M_c$ for both testbeds as 12 and 8, under which the localization mean errors reach minimum. It is fair to state that for the larger and multipath-richer room, a greater number of RP candidates should be required in order to well perform the position determination.

\subsubsection{Localization Accuracy in Both Testbeds}
In this part, by using the same parameters for all competing IPSs, we then move forward to evaluate the localization performance and present numerical results with relevant discussions. 

\begin{figure}[!t]
	\centering
	\includegraphics[width=1\linewidth]{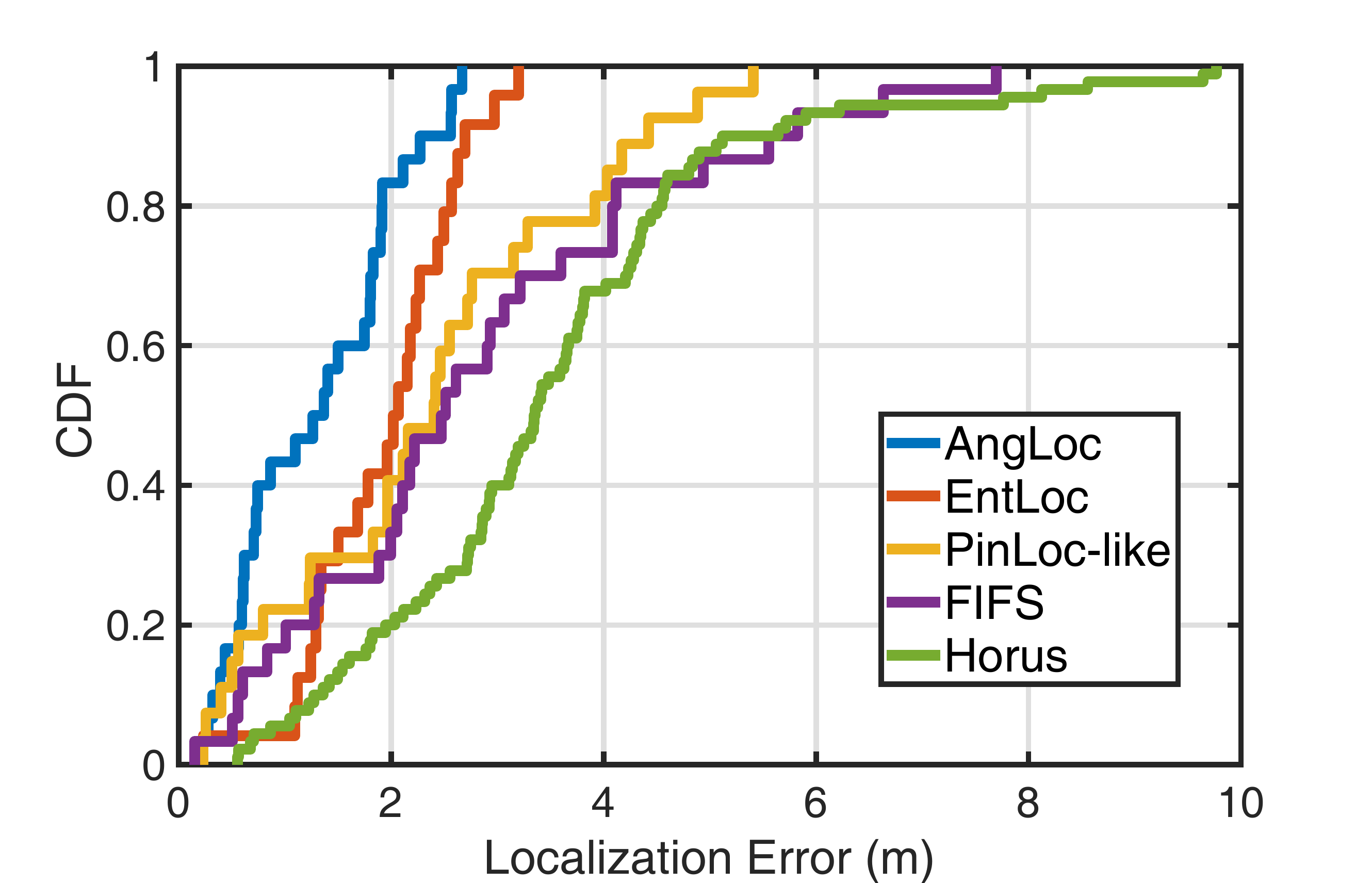}
	\caption{Localization accuracy for the laboratory (Testbed $\mypound 1$)}
	\label{cdfall1}
\end{figure}

\begin{figure}[!t]
	\centering
	\includegraphics[width=1\linewidth]{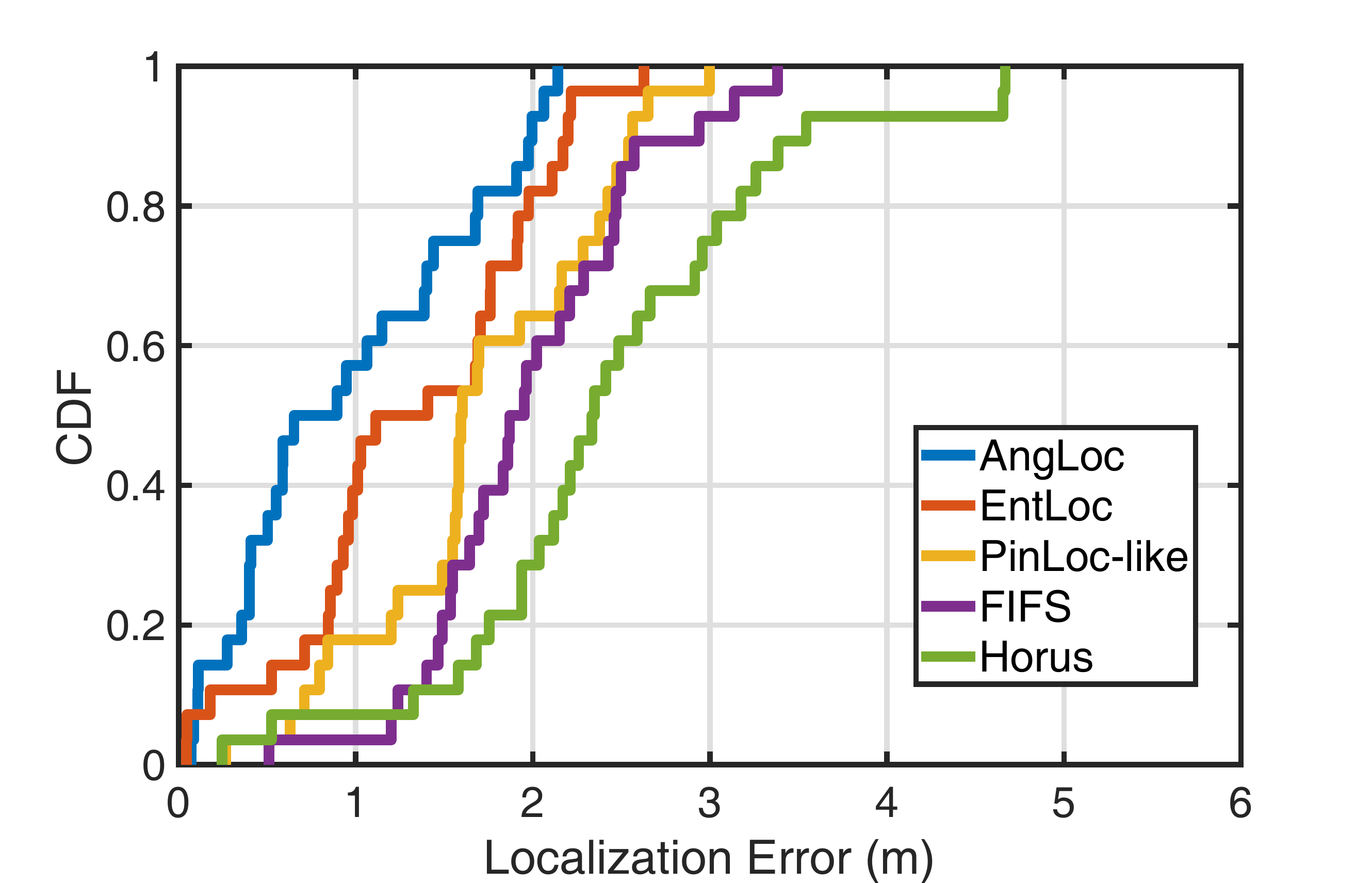}
	\caption{Localization accuracy for the classroom (Testbed $\mypound 2$)}
	\label{cdfall2}
\end{figure}

By virtue of cumulative distribution function (CDF), we first evaluate the localization accuracy of our proposed AngLoc system in comparison with the state-of-the-art. As can be observed in Fig. \ref{cdfall1}, for the laboratory environment, our proposed system is able to achieve the 90th percentile error of 2.27m, which outperforms EntLoc, PinLoc-like, FIFS and Horus systems with the same error level of 2.69m, 4.15m, 5.56m and 5.64m, respectively. Similarly, in the classroom scenario, we can notice in Fig. \ref{cdfall2} that AngLoc still precedes other rivals in terms of 90th percentile error. Concretely, it can ensure $90\%$ of test locations have a positioning error under 1.99m, surpassing EntLoc, PinLoc-like, FIFS and Horus systems with the same error percentage of $82.1\%$, $64.3\%$, $57.1\%$ and $28.6\%$, respectively. 

%\begin{figure*}[!t]
%	\begin{minipage}[t]{0.5\linewidth} 
%		\centering
%		\includegraphics[width=1\linewidth]{figures/cdf_compareall_1.pdf}
%		\caption{Localization accuracy for the laboratory (Testbed $\mypound 1$)} \label{cdfall1}
%	\end{minipage}
%	\begin{minipage}[t]{0.5\linewidth} 
%		\centering
%		\includegraphics[width=1\linewidth]{figures/cdf_compareall_2.pdf}
%		\caption{Localization accuracy for the classroom (Testbed $\mypound 2$)} \label{cdfall2}
%	\end{minipage}
%\end{figure*}

\begin{figure}[!t]
	\centering
	\includegraphics[width=1\linewidth]{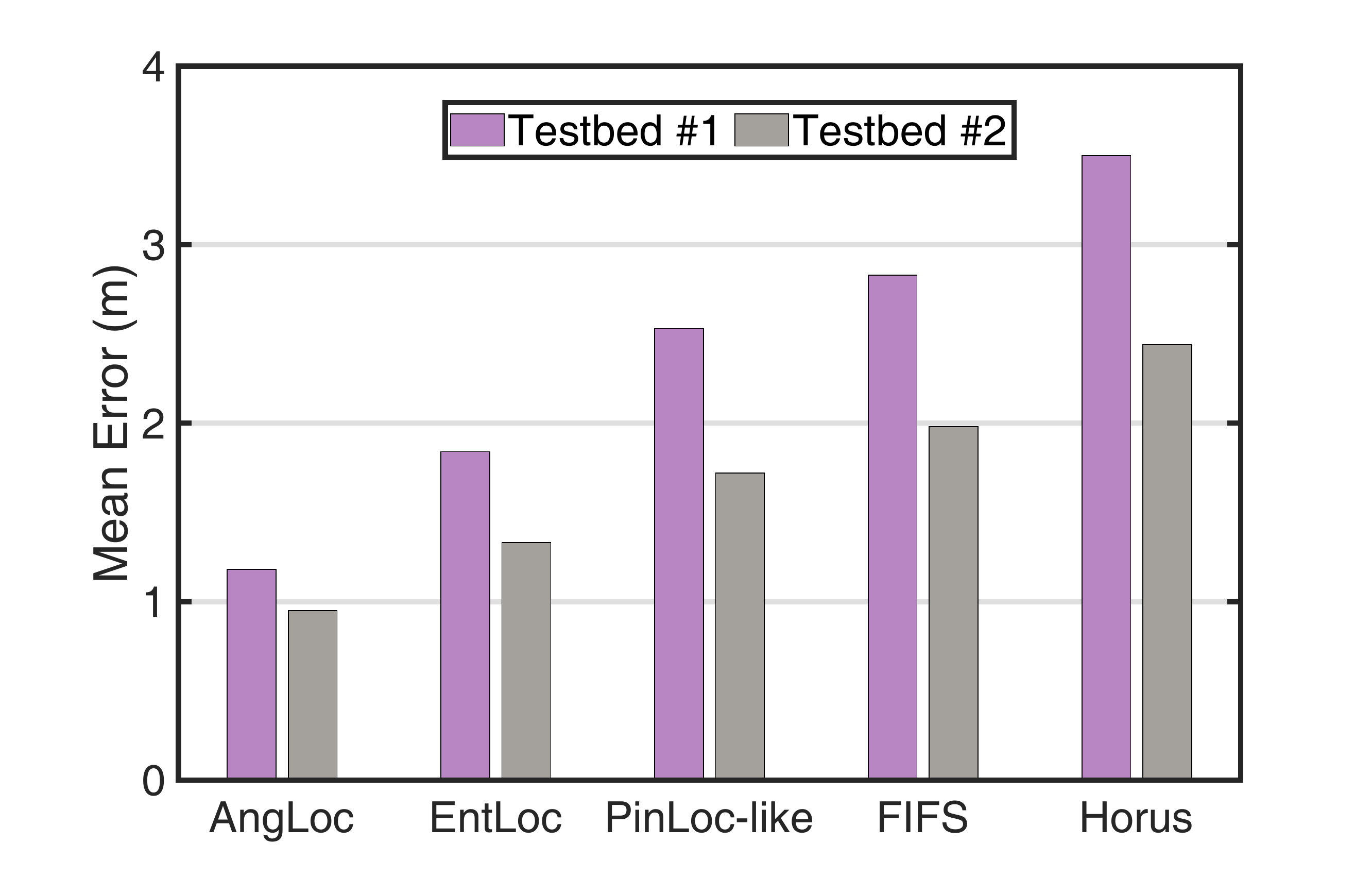}
	\caption{Bar plot of localization mean error comparison for both testbeds.}
	\label{meanbar}
\end{figure}

Moreover, since the classroom environment is relatively smaller and exposes more LoS radio propagation than the laboratory testbed, a superior localization performance is expected under the same conditions. We further display the mean error bar plot in Fig. \ref{meanbar} to provide an intuitive comparison within the five candidate systems. As expected, our proposed AngLoc system shows a mean error of 1.18m in the lab and 0.95m in the classroom, which even achieves the decimeter-level localization accuracy, outperforming other counterparts in both testbeds. Meanwhile, for all the competing IPSs, we can also observe that the mean error performance in the classroom is generally better than that in the laboratory scenario, which further validates our previous assumption.

In order to provide an in-depth and comprehensive comparison for these localization systems, we also enumerate the respective maximum error (Max. err.), minimum error (Min. err.), mean error (Mean err.) and the 90th percentile accuracy (Acc. at $90\%$) in Table \ref{errtable1} and Table \ref{errtable2} for the laboratory and classroom, respectively. As can be observed, apart from the minimum error, our AngLoc system broadly dominates the general accuracy evaluation for the maximum error, mean error and 90th percentile accuracy. When it comes to the particular minimum error, AngLoc only falls behind FIFS and EntLoc with 0.01m and 0.03m in the respective testbed $\mypound 1$ and $\mypound 2$, which can be reasonably neglected in both realistic indoor environments.

\begin{table}[t!]
	\centering
	\caption{\textbf{Localization accuracy for the lab scenario}}
	\label{errtable1}
	\bgroup
	\def\arraystretch{1.5}%  1 is the default, change whatever you need; change table gap (Luan)
	\begin{tabular}{l c c c c}
		\toprule 
		Methods & Max. err. & Min. err. & Mean err. & Acc. at 90\% \\ 
		\midrule 
		AngLoc  &\cellcolor{gray!25}{{2.67m}}&0.16m&\cellcolor{gray!25}{{1.18m}}&\cellcolor{gray!25}{{2.27m}}\\
		%\hline
		EntLoc      & 3.20m          & 0.23m         & 1.84m          & 2.69m          \\ 
		%\hline 
		PinLoc-like	& 5.85m          & 0.46m         & 2.53m          & 4.15m          \\ 
		%\hline 
		FIFS	    & 7.70m          &\cellcolor{gray!25}{{0.15m}}& 2.83m    & 5.56m   \\ 
		%\hline 
		Horus	    & 9.77m          & 0.55m         & 3.50m          & 5.64m          \\ 
		\bottomrule 
	\end{tabular}
	\egroup
\end{table}

\begin{table}[t!]
	\centering
	\caption{\textbf{Localization accuracy for the classroom scenario}}
	\label{errtable2}
	\bgroup
	\def\arraystretch{1.5}%  1 is the default, change whatever you need; change table gap (Luan)
	\begin{tabular}{l c c c c}
		\toprule
		Methods     & Max. err.      & Min. err.     & Mean err.      & Acc. at 90\%   \\
		\midrule 
		AngLoc      
		&\cellcolor{gray!25}{{2.14m}}&0.07m&\cellcolor{gray!25}{{0.95m}}&\cellcolor{gray!25}{{1.99m}}\\ 
		%\hline 
		EntLoc      & 2.62m     & \cellcolor{gray!25}{{0.04m}}& 1.33m          & 2.20m  \\ 
		%\hhline{~~-~~}
		%\hline 
		PinLoc-like	& 2.99m          & 0.27m         & 1.72m          & 2.56m          \\ 
		%\hline 
		FIFS	    & 3.38m          & 0.51m         & 1.98m          & 2.94m          \\ 
		%\hline 
		Horus	    & 4.67m          & 0.24m         & 2.44m          & 3.54m          \\ 
		\bottomrule
	\end{tabular}
	\egroup
\end{table}

%\subsection{Effect of AoA}
%Given that we have already AR entropy in \cite{luan2019entropy}, in this section, we focus on the fingerprinting performance effect imposed by the AoA fingerprint. 

\subsubsection{AoA Estimation Accuracy in LoS Condition}
In comparison with our previous EntLoc system, the most productive advancement for AngLoc is that CSI phase based AoA information is organically combined to facilitate the improvement of localization performance. Since one of SpotFi's key insights is to identify the direct path AoA for geometric mapping, even in strong NLoS case, it still needs multiple APs to achieve this through a likelihood scheme. In contrast, the inherent difference of our AngLoc is that the physical direct path AoA is not necessary for fingerprinting as long as the test target's AoA reading (i.e. the first arrival path) is similar with those of its neighboring RP locations. In order to create a fair competition, we only compare the AoA estimation errors with SpotFi under the LoS condition. For the NLoS scenario, we design a different evaluation mechanism for the comparison purpose, which will be discussed in the next part. 

\begin{figure}[!t]
	\centering
	\includegraphics[width=1\linewidth]{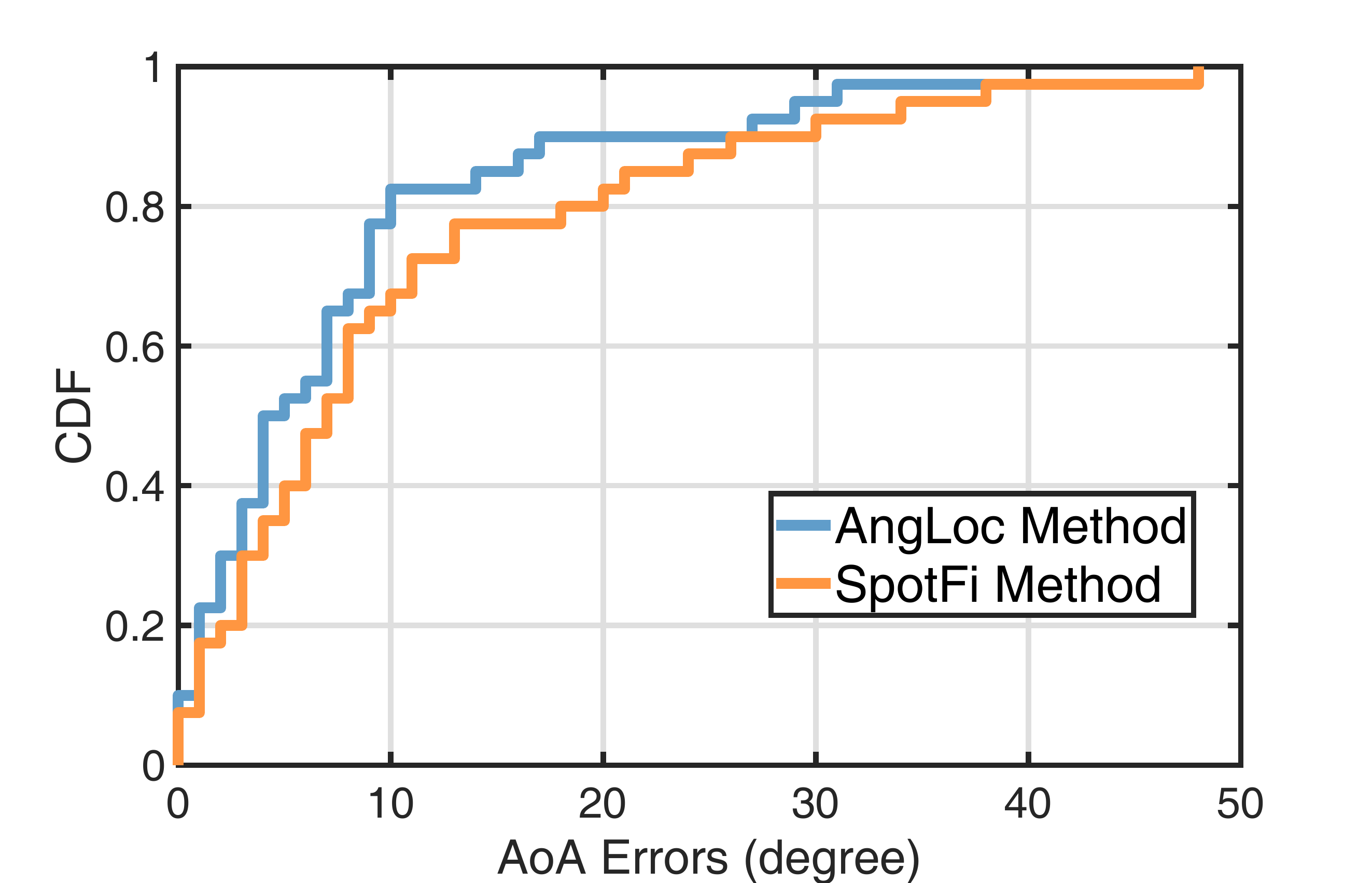}
	\caption{AoA estimation errors in LoS condition}
	\label{aoaerr}
\end{figure}

\begin{figure}[!t]
	\centering
	\includegraphics[width=1\linewidth]{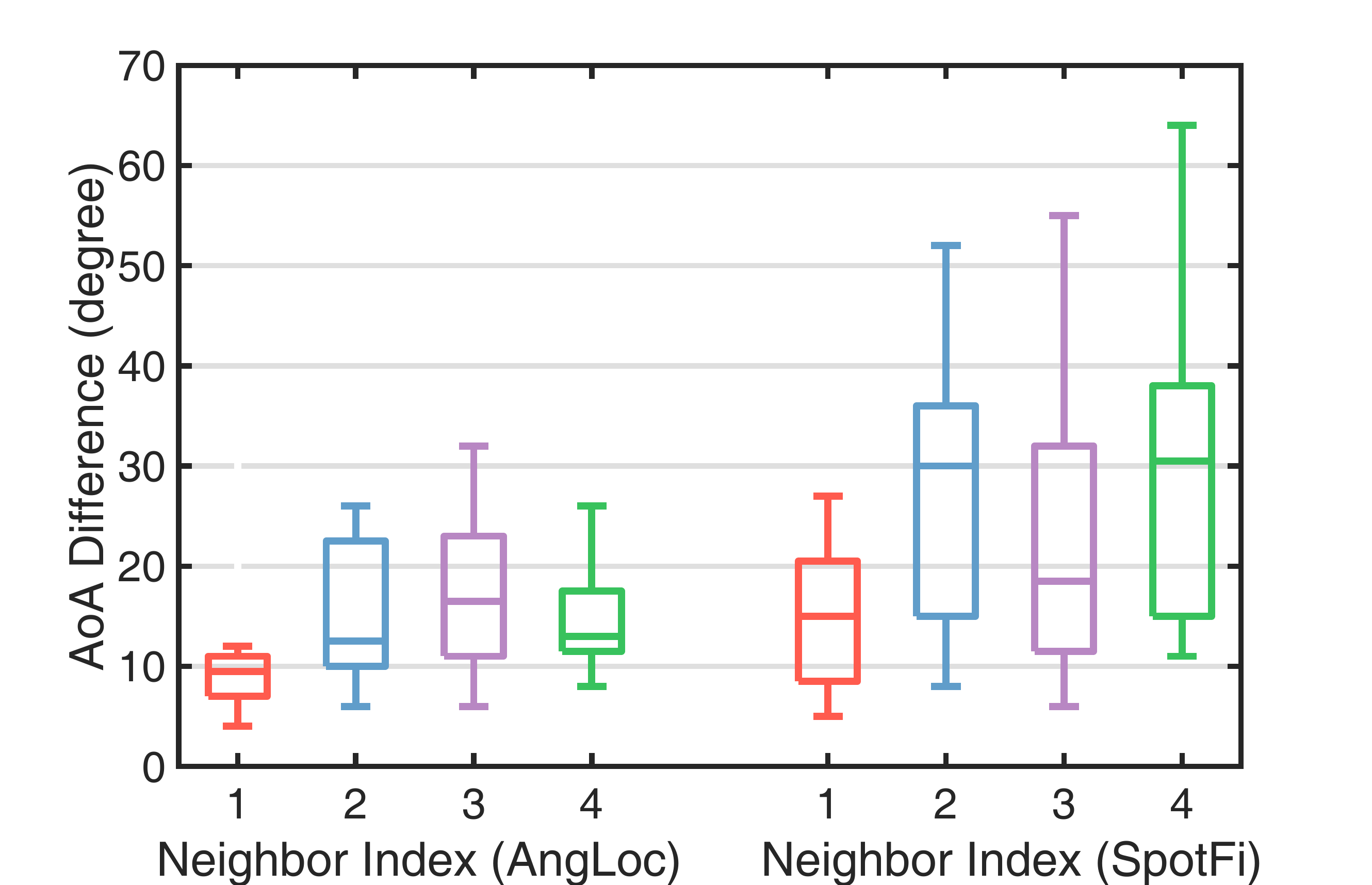}
	\caption{The box plot for AoA differences between 20 test locations and their corresponding 4 neighboring RP locations.}
	\label{aoabox}
\end{figure}

In practice, for the LoS classroom illustrated in Fig. \ref{fplan_cr}, by applying our AngLoc's enhanced AoA estimation approach as well as the SpotFi's method, we record the AoA readings at all 40 RP locations and compare them with their corresponding ground truth. Here it is worth mentioning that after obtaining several AoA-ToF estimates (clusters) from multiple CSI packets, SpotFi declares the direct path AoA from the cluster with the highest likelihood value. For a fair comparison, we modify the last part of SpotFi to determine the AoA from the first arrival path, which is exactly what we adopt in the AngLoc system. As shown in Fig. \ref{aoaerr}, our AngLoc's AoA estimation method can yield the 90th percentile error of 17 degrees, outperforming SpotFi's 26 degrees error in the LoS condition. The total gain of nearly 10 degrees validates the superior performance of our super-resolution JADE-MUSIC algorithm.

\subsubsection{Impact of AoA Proximity in NLoS Condition}
For NLoS environment such as the laboratory shown in Fig. \ref{fplan_lab}, we design and implement a dedicated experiment to manifest the AoA based fingerprinting feasibility of our AngLoc system. Specifically, we first choose 20 test locations which are in the obvious NLoS conditions from the transmitter. Each of them is surrounded by four predefined RPs. After acquiring the AoA estimates from all these test positions and their neighboring RP locations, we then calculate the AoA differences between each test location and its corresponding four RPs in the vicinity. As displayed in Fig. \ref{aoabox}, compared with SpotFi's method, the box plot shows that our AngLoc's AoA estimation method is capable of deriving the overall lower level of AoA differences with the four RP neighbors, which nicely indicates the similar AoA estimations around the neighboring locations. This advantage further promotes our AngLoc-derived AoA to be a well-qualified position fingerprint for the accurate indoor location determination.

%\begin{figure}[!t]
%\centering
%\subfloat[AngLoc Method]{\includegraphics[width=0.8\linewidth,height=1.3in]{figures/aoa_box_nlos1.pdf}
%\label{aoabox1}}
%\hfil
%\subfloat[SpotFi Method]{\includegraphics[width=0.8\linewidth,height=1.3in]{figures/aoa_box_nlos2.pdf}
%\label{aoabox2}} 
%\caption{The box plot for AoA differences between 20 test locations and their corresponding 4 neighboring RP locations.}
%\end{figure}

%%%%%%%%%%%%%%%%%%%%%%%%%%%%%%%%%%%%%%%%%%%%%%%%%%%%%%%%%%%%%%%%%%%%%%%%%%%%%%%%%%%%%%%%%%%%%%%%%%%
\section{Discussion}\label{sec:discussion}
In this section, we discuss several unsolved issues in this paper and propose some possible solutions, which could further enhance the performance of our proposed localization system.

\subsection{Device Orientation Calibration}
During the fingerprint site survey, we frequently move the HummingBoard embedded antenna stand among all the RP locations to collect CSI data. In principle, to achieve preferable AoA estimations, one should always maintain the same device orientation when moving the antenna stand from place to place. Otherwise, it may incur additional manual operational error for AoA estimates. To address this potential problem, the commodity smart robot can be leveraged in our future work, by which we can plan the moving path in advance and fix the device orientation automatically, thus further improving AoA based location fingerprinting accuracy.

%\subsection{Geolocation for Multiple Targets}
%To advance our AngLoc prototype to a broader range of the commercial applications, how to simultaneous multi-target positioning is still challenging. A promising 

\subsection{Alternative Hardware Implementation}
As aforementioned in Section \ref{sec:experiment}, our entire experimental framework is established on the basis of commodity wireless IWL 5300 NIC chipset, which provides IEEE 802.11n CSI in a format of 30 subcarrier groups for both 20 MHz and 40 MHz bandwidth. In practice, this sets the limitation for some CSI based applications which demand higher resolution of CSI subcarriers. Such examples include human activity recognition \cite{yang2018device}, indoor distance ranging \cite{zhu2018pi} and so forth. Recently, some other CSI tools like Atheros CSI tool \cite{zhuo2017perceiving} is getting prevalent in the academic domain due to its non-grouping and non-compressed CSI reporting. Unlike Intel's 5300 NIC, the Qualcomm Atheros NIC chipset is able to report CSI value for each subcarrier, i.e., 56 subcarriers for 20MHz channel and 114 subcarriers for a 40MHz channel. Furthermore, it can also display detailed payload records and retrieve rich status information about the received packet. These additional CSI information can be of great value to help further enhance the localization performance of our AngLoc fingerprinting system.

%%%%%%%%%%%%%%%%%%%%%%%%%%%%%%%%%%%%%%%%%%%%%%%%%%%%%%%%%%%%%%%%%%%%%%%%%%%%%%%%%%%%%%%%%%%%%%%%%%%
\section{Conclusion}\label{sec:conclusion}
In this paper, we presented AngLoc, an AoA-aware probabilistic indoor location fingerprinting system using CSI information. In AngLoc, a tap filtering scheme was first proposed to remove the noisy component in raw CSI measurements. Meanwhile, for achieving accurate AoA estimation, we employed several phase calibration techniques to further compensate CSI phase errors. In the offline phase, we adopted AR modeling entropy as the amplitude based fingerprint since it captures the most informative statistical information of CSI amplitude while maintaining a simple structure. In addition, an enhanced JADE-MUSIC algorithm was leveraged to derive AoA estimates as the CSI phase based fingerprint. A robust radio map containing both CSI amplitude and phase information is then readily constructed. In the online phase, for a mobile target, we first narrowed down the candidate RP locations by finding RPs with the smallest AR entropy differences. A novel bivariate kernel regression method was then adopted to precisely infer the target's location. In comparison with our previous EntLoc system, experimental results from the lightweight HummingBoard device showed a superior localization performance of our proposed AngLoc system with an average accuracy improvement of $35.9\%$ and $28.6\%$ in both laboratory and classroom testbeds. Additionally, we also examined the impacts of several parameters on AngLoc's performance in different indoor scenarios, which empowers us with deepening insights to efficiently and productively conduct our indoor location fingerprinting.

% or


%\appendices

%\section{}

%\section*{Acknowledgment}
%The authors would like to thank the anonymous reviewers for their valuable comments and suggestions.
% The authors would like to thank Mr. Christophe Alexandre for his valuable assistance in the domain of Linux development as well as hardware implementation.
\ifCLASSOPTIONcaptionsoff
  \newpage
\fi



\bibliographystyle{IEEEtran}
\bibliography{refluan}

\begin{thebibliography}{10}
\providecommand{\url}[1]{#1}
\csname url@samestyle\endcsname
\providecommand{\newblock}{\relax}
\providecommand{\bibinfo}[2]{#2}
\providecommand{\BIBentrySTDinterwordspacing}{\spaceskip=0pt\relax}
\providecommand{\BIBentryALTinterwordstretchfactor}{4}
\providecommand{\BIBentryALTinterwordspacing}{\spaceskip=\fontdimen2\font plus
\BIBentryALTinterwordstretchfactor\fontdimen3\font minus
  \fontdimen4\font\relax}
\providecommand{\BIBforeignlanguage}[2]{{%
\expandafter\ifx\csname l@#1\endcsname\relax
\typeout{** WARNING: IEEEtran.bst: No hyphenation pattern has been}%
\typeout{** loaded for the language `#1'. Using the pattern for}%
\typeout{** the default language instead.}%
\else
\language=\csname l@#1\endcsname
\fi
#2}}
\providecommand{\BIBdecl}{\relax}
\BIBdecl

\bibitem{kuutti2018survey}
S.~Kuutti, S.~Fallah, K.~Katsaros, M.~Dianati, F.~Mccullough, and
  A.~Mouzakitis, ``{A Survey of the State-of-the-Art Localization Techniques
  and Their Potentials for Autonomous Vehicle Applications},'' \emph{IEEE
  Internet Things J.}, vol.~5, no.~2, pp. 829--846, 2018.

\bibitem{zafari2019survey}
F.~Zafari, A.~Gkelias, and K.~K. Leung, ``{A Survey of Indoor Localization
  Systems and Technologies},'' \emph{IEEE Commun. Surveys Tuts.}, 2019.

\bibitem{he2015wi}
S.~He and S.-H.~G. Chan, ``{Wi-Fi Fingerprint-based Indoor Positioning: Recent
  Advances and Comparisons},'' \emph{IEEE Commun. Surveys Tuts.}, vol.~18,
  no.~1, pp. 466--490, 2015.

\bibitem{dardari2015indoor}
D.~Dardari, P.~Closas, and P.~M. Djuri{\'c}, ``{Indoor Tracking: Theory,
  Methods, and Technologies},'' \emph{IEEE Trans. Veh. Technol.}, vol.~64,
  no.~4, pp. 1263--1278, 2015.

\bibitem{liu2013face}
S.~Liu, Y.~Jiang, and A.~Striegel, ``{Face-to-Face Proximity Estimation using
  Bluetooth on Smartphones},'' \emph{IEEE Trans. Mobile Comput.}, vol.~13,
  no.~4, pp. 811--823, 2013.

\bibitem{ni2003landmarc}
L.~M. Ni, Y.~Liu, Y.~C. Lau, and A.~P. Patil, ``{LANDMARC: Indoor Location
  Sensing using Active RFID},'' in \emph{Proc. IEEE PerCom}, 2003, pp.
  407--415.

\bibitem{ward1997new}
A.~Ward, A.~Jones, and A.~Hopper, ``{A New Location Technique for the Active
  Office},'' \emph{IEEE Personal Commun.}, vol.~4, no.~5, pp. 42--47, 1997.

\bibitem{want1992active}
R.~Want, A.~Hopper, V.~Falcao, and J.~Gibbons, ``{The Active Badge Location
  System},'' \emph{ACM Trans. Inf. Syst.}, vol.~10, no.~1, pp. 91--102, 1992.

\bibitem{pathak2015visible}
P.~H. Pathak, X.~Feng, P.~Hu, and P.~Mohapatra, ``{Visible Light Communication,
  Networking, and Sensing: A Survey, Potential and Challenges},'' \emph{IEEE
  Commun. Surveys Tuts.}, vol.~17, no.~4, pp. 2047--2077, 2015.

\bibitem{yang2013rssi}
Z.~Yang, Z.~Zhou, and Y.~Liu, ``{From RSSI to CSI: Indoor Localization via
  Channel Response},'' \emph{ACM Comput. Surv.}, vol.~46, no.~2, p.~25, 2013.

\bibitem{tadayon2019decimeter}
N.~Tadayon, M.~T. Rahman, S.~Han, S.~Valaee, and W.~Yu, ``{Decimeter Ranging
  with Channel State Information},'' \emph{IEEE Trans. Wireless Commun.}, 2019.

\bibitem{kotaru2015spotfi}
M.~Kotaru, K.~Joshi, D.~Bharadia, and S.~Katti, ``{SpotFi: Decimeter Level
  Localization using WiFi},'' \emph{ACM SIGCOMM Comput. Commun. Rev.}, vol.~45,
  no.~4, pp. 269--282, 2015.

\bibitem{wu2012csi}
K.~Wu, J.~Xiao, Y.~Yi, D.~Chen, X.~Luo, and L.~M. Ni, ``{CSI-based Indoor
  Localization},'' \emph{IEEE Trans. Parallel Distrib. Syst.}, vol.~24, no.~7,
  pp. 1300--1309, 2012.

\bibitem{alsindi2014empirical}
N.~Alsindi, Z.~Chaloupka, N.~AlKhanbashi, and J.~Aweya, ``{An Empirical
  Evaluation of A Probabilistic RF Signature for WLAN Location
  Fingerprinting},'' \emph{IEEE Trans. Wireless Commun.}, vol.~13, no.~6, pp.
  3257--3268, 2014.

\bibitem{youssef2005horus}
M.~Youssef and A.~Agrawala, ``{The Horus WLAN Location Determination System},''
  in \emph{Proc. ACM MobiSys}, 2005, pp. 205--218.

\bibitem{xiao2012fifs}
J.~Xiao, K.~Wu, Y.~Yi, and L.~M. Ni, ``{FIFS: Fine-grained Indoor
  Fingerprinting System},'' in \emph{Proc. IEEE ICCCN}, 2012, pp. 1--7.

\bibitem{sen2012you}
S.~Sen, B.~Radunovic, R.~R. Choudhury, and T.~Minka, ``{You are Facing the Mona
  Lisa: Spot Localization using PHY Layer Information},'' in \emph{Proc. ACM
  MobiSys}, 2012, pp. 183--196.

\bibitem{zhou2013omnidirectional}
Z.~Zhou, Z.~Yang, C.~Wu, L.~Shangguan, and Y.~Liu, ``{Omnidirectional Coverage
  for Device-Free Passive Human Detection},'' \emph{IEEE Trans. Parallel
  Distrib. Syst.}, vol.~25, no.~7, pp. 1819--1829, 2013.

\bibitem{mirowski2011kl}
P.~Mirowski, H.~Steck, P.~Whiting, R.~Palaniappan, M.~MacDonald, and T.~K. Ho,
  ``{KL-Divergence Kernel Regression for Non-Gaussian Fingerprint based
  Localization},'' in \emph{Proc. IEEE IPIN}, 2011, pp. 1--10.

\bibitem{chen2018probabilistic}
L.~Chen, I.~Ahriz, D.~Le~Ruyet, and H.~Sun, ``{Probabilistic Indoor Position
  Determination via Channel Impulse Response},'' in \emph{Proc. IEEE PIMRC},
  2018, pp. 829--834.

\bibitem{bercher2000estimating}
J.-F. Bercher and C.~Vignat, ``{Estimating the Entropy of a Signal with
  Applications},'' \emph{IEEE Trans. Signal Process.}, vol.~48, no.~6, pp.
  1687--1694, 2000.

\bibitem{kay1998model}
S.~Kay, ``{Model-based Probability Density Function Estimation},'' \emph{IEEE
  Signal Process. Lett.}, vol.~5, no.~12, pp. 318--320, 1998.

\bibitem{luan2019entropy}
L.~{Chen}, I.~{Ahriz}, and D.~{Le Ruyet}, ``{CSI-based Probabilistic Indoor
  Position Determination: An Entropy Solution},'' \emph{IEEE Access}, vol.~7,
  pp. 170\,048--170\,061, 2019.

\bibitem{molisch2012wireless}
A.~F. Molisch, \emph{{Wireless Communications}}.\hskip 1em plus 0.5em minus
  0.4em\relax John Wiley \& Sons, 2012.

\bibitem{vanderveen1997joint}
M.~C. Vanderveen, C.~B. Papadias, and A.~Paulraj, ``{Joint Angle and Delay
  Estimation (JADE) for Multipath Signals Arriving at an Antenna Array},''
  \emph{IEEE Commun. Lett.}, vol.~1, no.~1, pp. 12--14, 1997.

\bibitem{wu2012fila}
K.~Wu, J.~Xiao, Y.~Yi, M.~Gao, and L.~M. Ni, ``{FILA: Fine-grained Indoor
  Localization},'' in \emph{Proc. IEEE INFOCOM}, 2012, pp. 2210--2218.

\bibitem{xiong2013arraytrack}
J.~Xiong and K.~Jamieson, ``{ArrayTrack: A Fine-Grained Indoor Location
  System},'' in \emph{Proc. USENIX NSDI 13}, 2013, pp. 71--84.

\bibitem{vasisht2016decimeter}
D.~Vasisht, S.~Kumar, and D.~Katabi, ``{Decimeter-Level Localization with a
  Single WiFi Access Point},'' in \emph{Proc. USENIX NSDI 16}, 2016, pp.
  165--178.

\bibitem{bahl2000radar}
P.~{Bahl} and V.~N. {Padmanabhan}, ``{RADAR: An In-Building RF-based User
  Location and Tracking System},'' in \emph{Proc. IEEE INFOCOM}, vol.~2, 2000,
  pp. 775--784.

\bibitem{wang2016csi}
X.~Wang, L.~Gao, S.~Mao, and S.~Pandey, ``{CSI-based Fingerprinting for Indoor
  Localization: A Deep Learning Approach},'' \emph{IEEE Trans. Veh. Technol.},
  vol.~66, no.~1, pp. 763--776, 2016.

\bibitem{Halperin_csitool}
D.~Halperin, W.~Hu, A.~Sheth, and D.~Wetherall, ``{Tool Release: Gathering
  802.11n Traces with Channel State Information},'' \emph{ACM SIGCOMM Comput.
  Commun. Rev.}, vol.~41, no.~1, p.~53, Jan. 2011.

\bibitem{chen2016achieving}
C.~Chen, Y.~Chen, Y.~Han, H.-Q. Lai, and K.~R. Liu, ``{Achieving
  Centimeter-Accuracy Indoor Localization on WiFi Platforms: A Frequency
  Hopping Approach},'' \emph{IEEE Internet Things J.}, vol.~4, no.~1, pp.
  111--121, 2016.

\bibitem{zhou2015wifi}
Z.~Zhou, Z.~Yang, C.~Wu, L.~Shangguan, H.~Cai, Y.~Liu, and L.~M. Ni,
  ``{WiFi-based Indoor Line-of-Sight Identification},'' \emph{IEEE Trans.
  Wireless Commun.}, vol.~14, no.~11, pp. 6125--6136, 2015.

\bibitem{ma2019wifi}
Y.~Ma, G.~Zhou, and S.~Wang, ``{WiFi Sensing with Channel State Information: A
  Survey},'' \emph{ACM Comput. Surv.}, vol.~52, no.~3, p.~46, 2019.

\bibitem{zhuo2017perceiving}
Y.~Zhuo, H.~Zhu, H.~Xue, and S.~Chang, ``{Perceiving Accurate CSI Phases with
  Commodity WiFi Devices},'' in \emph{Proc. IEEE INFOCOM}, 2017, pp. 1--9.

\bibitem{ma2018signfi}
Y.~Ma, G.~Zhou, S.~Wang, H.~Zhao, and W.~Jung, ``{SignFi: Sign Language
  Recognition using WiFi},'' \emph{Proc. ACM Interact. Mob. Wearable Ubiquitous
  Technol.}, vol.~2, no.~1, p.~23, 2018.

\bibitem{kay1993fundamentals}
S.~M. Kay, \emph{{Fundamentals of Statistical Signal Processing}}.\hskip 1em
  plus 0.5em minus 0.4em\relax Prentice Hall PTR, 1993.

\bibitem{kay2005exponentially}
S.~Kay, ``{Exponentially Embedded Families-New Approaches to Model Order
  Estimation},'' \emph{IEEE Trans. Aerosp. Electron. Syst.}, vol.~41, no.~1,
  pp. 333--345, 2005.

\bibitem{schmidt1986multiple}
R.~Schmidt, ``{Multiple Emitter Location and Signal Parameter Estimation},''
  \emph{IEEE Trans. Antennas Propag.}, vol.~34, no.~3, pp. 276--280, 1986.

\bibitem{czink2004number}
N.~Czink, M.~Herdin, H.~{\"O}zcelik, and E.~Bonek, ``{Number of Multipath
  Clusters in Indoor MIMO Propagation Environments},'' \emph{Electron. Lett.},
  vol.~40, no.~23, pp. 1498--1499, 2004.

\bibitem{pillai1989forward}
S.~U. Pillai and B.~H. Kwon, ``{Forward/Backward Spatial Smoothing Techniques
  for Coherent Signal Identification},'' \emph{IEEE Trans. Acoust., Speech,
  Signal Process.}, vol.~37, no.~1, pp. 8--15, 1989.

\bibitem{krause1986taxicab}
E.~F. Krause, \emph{{Taxicab Geometry: An Adventure in Non-Euclidean
  Geometry}}.\hskip 1em plus 0.5em minus 0.4em\relax Courier Corporation, 1986.

\bibitem{cover2012elements}
T.~M. Cover and J.~A. Thomas, \emph{{Elements of Information Theory}}.\hskip
  1em plus 0.5em minus 0.4em\relax John Wiley \& Sons, 2012.

\bibitem{yang2018device}
J.~Yang, H.~Zou, H.~Jiang, and L.~Xie, ``{Device-free Occupant Activity Sensing
  using WiFi-enabled IoT Devices for Smart Homes},'' \emph{IEEE Internet Things
  J.}, vol.~5, no.~5, pp. 3991--4002, 2018.

\bibitem{zhu2018pi}
H.~Zhu, Y.~Zhuo, Q.~Liu, and S.~Chang, ``$\pi$-splicer: Perceiving accurate csi
  phases with commodity wifi devices,'' \emph{IEEE Trans. Mobile Comput.},
  vol.~17, no.~9, pp. 2155--2165, 2018.

\end{thebibliography}

\vfill


\end{document}